\documentclass[]{autart}

\usepackage{amsmath}
\usepackage{bbold}
\usepackage{enumerate}
\usepackage{cite}
\usepackage[dvipsnames]{xcolor}
\usepackage{amssymb}
\usepackage{graphicx}
\usepackage{algorithm}
\usepackage{algpseudocode}
\usepackage{bbm}
\usepackage{subfigure}
\usepackage{epsfig}
\usepackage{epstopdf}
\usepackage{makecell}
\usepackage{caption}
\usepackage{threeparttable}
\usepackage{stfloats}
\usepackage{extarrows}

\def\dref#1{(\ref{#1})}

\newtheorem{lemma}{Lemma}
\newtheorem{assumption}{Assumption}
\newtheorem{theorem}{Theorem}
\newtheorem{proposition}{Proposition}
\newtheorem{corollary}{Corollary}
\newtheorem{remark}{Remark}

\begin{document}

\begin{frontmatter}

\title{{Robust Data-Driven Kalman Filtering for Unknown Linear Systems using Maximum Likelihood Optimization} \thanksref{mytitle}}

\thanks[mytitle]{This work was partially supported by Swedish Research Council Distinguished Professor Grant 2017-01078 and a Knut and Alice Wallenberg Foundation Wallenberg Scholar Grant. }

\thanks{P. Duan, Y. Xing, and K.H. Johansson are also affiliated with Digital Futures, Stockholm, Sweden. }

\author[mymainaddress]{Peihu~Duan}\ead{peihu@kth.se},    % Add the
\author[mysecondaryaddress]{Tao~Liu}\ead{taoliu@eee.hku.hk},  % (ead) as shown
\author[mymainaddress]{Yu Xing}\ead{yuxing2@kth.se},
\author[mymainaddress]{Karl Henrik Johansson}\ead{kallej@kth.se}  % (ead) as shown

\address[mymainaddress]{School of Electrical Engineering and Computer Science, KTH Royal Institute of Technology, Stockholm, Sweden}

\address[mysecondaryaddress]{Department of Electrical and Electronic Engineering, The University of Hong Kong, Hong Kong SAR, China}

\maketitle
\begin{abstract}
{This paper investigates the state estimation problem for unknown linear systems  subject to both process and measurement noise. Based on a prior input-output trajectory sampled at a higher frequency and a prior state trajectory sampled at a lower frequency, we propose a novel robust data-driven Kalman filter (RDKF) that integrates model identification with state estimation for the unknown system. Specifically, the state estimation problem is formulated as a non-convex maximum likelihood optimization problem. Then, we slightly modify the optimization problem  to get a problem solvable with a recursive algorithm. Based on the optimal solution to this new problem, the RDKF is designed, which can estimate the state of a given but unknown state-space model. The performance gap between the RDKF and the optimal Kalman filter based on known system matrices is quantified through a sample complexity bound. In particular, when the number of the pre-collected states tends to infinity, this gap converges to zero. Finally, the effectiveness of the theoretical results is illustrated by numerical simulations.}
\end{abstract}


\end{frontmatter}

\section{Introduction}\label{s1}
Due to its ability to estimate states of dynamic systems, Kalman filtering has attracted tremendous attention since its inception in 1960, which has been widely applied in practice \cite{kalman1960new1,Auger2013}. However, the effectiveness of Kalman filters is dependent on prior knowledge of system dynamics that may be unavailable in some practical implementations \cite{anderson2005optimal,netto2018robust}. To tackle this issue, some research efforts have been devoted to learning filters from pre-collected system trajecotries, as described below.

%Nevertheless, some tricky issues in the design of data-driven filters remain open, including 1) determining a given but unknown state-space model using pre-collected system data for the filter design; 2) preserving advantages, such as optimality and recursive structure of the model-based Kalman filter in data-driven filters.

% \subsection{Related Works} \label{s1.1}
Learning state estimators for unknown systems from pre-collected system data has been a longstanding topic in the control society \cite{hou2013model,markovsky2021behavioral}. According to designing criteria, there are two paradigms of data-driven state estimation: {\it indirect data-driven state estimation}, also referred to as system identification-based state estimation \cite{tsiamis2020sample,alanwar2022data,Guy2022}, and {\it direct data-driven state estimation} \cite{mehra1970identification,shafieezadeh2018wasserstein,liu2023learning,Wolff2024}.
Indirect data-driven state estimation identifies a state-space model using the pre-collected system data following classical system identification approaches \cite{ljung1998system}, and then designs a state estimator based on the identified model. State-space linear models \cite{tsiamis2020sample,alanwar2022data} and neural networks (NNs) \cite{Guy2022} are commonly adopted. For example, Tsiamis  et al. \cite{tsiamis2020sample} adopted a subspace identification approach to identify a state-space linear model and later proposed both a certainty equivalent and a robust Kalman filter.
%Later, Alanwar et al. \cite{alanwar2022data} proposed a data-driven set-based estimator based on system identification, which could determine the specific state by utilizing prior state trajectory.
Revach et al. \cite{Guy2022} trained an NN model to describe a system before designing a Kalman filter. Direct data-driven state estimation designs a state estimator directly from system data with no intermediate system identification step. This method is based on different principles, such as adaptive control theory \cite{mehra1970identification,shafieezadeh2018wasserstein} and Willems' fundamental lemma \cite{liu2023learning,Wolff2024}. The pioneering idea of direct data-driven state estimation can be traced back to adaptive filtering for cases with unknown noise covariances  \cite{mehra1970identification,shafieezadeh2018wasserstein}. Recently, {Willems' fundamental lemma \cite[Theorem 1]{willems2005note}, which provides a sufficient condition under which an input-output trajectory of a linear time-invariant system can be recovered by a measured input-output trajectory,} contributes to several essential results on direct data-driven filtering  \cite{liu2023learning,Wolff2024}. For example, Liu et al. \cite{liu2023learning} and Wolff et al. \cite{Wolff2024} leveraged this lemma to design an explicit observer and an implicit moving horizon estimator for a state-space linear system with unknown system matrices, respectively. % In the above Willems' fundamental lemma-based state estimation methods, a continuous input-state-output trajectory is needed.

{The aforementioned data-driven state estimation methods rely on a prior input-state-output trajectory, where the input, state, and output should be sampled at the same frequency. However, in many practical scenarios, the state sampling frequency is often much lower than that of the input-output trajectory, as measuring the state may require additional sampling strategies and extended sampling times, as illustrated in the motivating example in Section \ref{s2}. In this case, an unresolved issue is how to leverage a prior input-output trajectory and a lower-frequency sampled state trajectory for online state estimation of an unknown system.} Moreover, the impact of data noise on the filtering performance is not yet fully understood in the literature. For example, the specific relationship between the magnitude of data noise and the filtering performance has not been established. Altogether, designing a filtering method that can estimate the state of a given state-space model with unknown system matrices has not well addressed, let along conducting an in-depth analysis of the filtering performance.

% \subsection{Contributions} \label{s1.3}
Motivated by the above findings, this paper investigates the state estimation problem for a linear system with a pre-defined state but unknown system matrices of the corresponding state-space model. {For this system, we pre-collect an input-output trajectory at a higher frequency and a state trajectory at a lower frequency.} This paper formulates the modeling and filtering problem for the unknown linear system as a unified maximum likelihood (ML) optimization problem, the solution to which generates a novel robust data-driven Kalman filter (RDKF). In comparison to the literature, this paper possesses several special features as follows:
\begin{enumerate}
  \item The RDKF is developed based on a prior input-output trajectory and a lower-frequency sampled state trajectory, providing an alternative for state estimation in practical scenarios where sampling the state at the same frequency as the input-output trajectory is impractical. 

  \item A feasibility condition for the RDKF is established, demonstrating that the RDKF is feasible when the prior system trajectory is sufficiently long, and providing a specific requirement on the trajectory length (\textbf{Theorem \ref{thm4}}).

  % Moreover, for the data collection, the proposed RDKF relies only on noisy partial states (i.e.,outputs) rather than accurate full states needed in \cite{verhaegen2007filtering,revach2021kalmannet,Guy2022,netto2018robust}.

  \item The filtering performance of the RDKF is ensured. Particularly, a sample-complexity bound is derived for the performance gap between the RDKF and the Kalman filter based on known system matrices, which also quantitatively reveals the impact of data noise on the performance (\textbf{Theorems \ref{thm5}} and \textbf{\ref{thm7}}).

  \item The RDKF is further generalized for cases with only a prior input-output trajectory. In this case, the state estimate corresponds to a balanced realization of the system, which can be applied to control tasks such as LQG control (\textbf{Corollary \ref{corlast}}). 

\end{enumerate}

The remainder of this paper is organized as follows. Section \ref{s2} presents the problem formulation. Section \ref{s3} introduces a novel RDKF algorithm. Section \ref{s4} analyzes the necessary informativity of the pre-collected data required for performing the RDKF. Section \ref{s5} evaluates the filtering performance of the RDKF. The theoretical results are illustrated in Section \ref{s7}. Section \ref{s8} concludes the paper.

 \noindent
{\it Notations}: Let $\mathbb{R}^{n}$ denote the real coordinate space of dimension $n$. Let $\otimes$ denote the Kronecker product. Let $I_n$ denote the $n$-order identity matrix. Let $0$ denote a scalar, vector, or matrix of an appropriate dimension with all elements being zero. Let $\mathbb{N}^{+}$ be the set of positive integers and $\mathbb{N} = \mathbb{N}^{+} \cup  0$. For any given vector $\mu$ and positive definite matrix $\Sigma$ with appropriate dimensions, let $\mathcal{N}(\mu,\ \Sigma)$ denote Gaussian distribution with mean $\mu$ and covariance $\Sigma$. For any positive function $f$, let $\text{ln} f$ denote its natural logarithm. For any matrix $S$, $S(m_1:m_2;n_1:n_2)$ denotes the block matrix in $S$ with elements $S_{ij}$, $ m_1 \leq i \leq m_2$, $n_1 \leq j \leq n_2$; $S^{\dagger}$ denotes its right/left inverse if it has full row/column rank. For a square matrix $S$, $|S|$ denotes its determinant; $\lambda(S)$ denotes the set of its eigenvalues; $\lambda_{\text{max}}(S)$/$\lambda_{\text{min}}(S)$ denotes its maximum/minimum eigenvalue if $S$ is positive definite. % ; and $S>0$ ($S \ge 0$) denotes $S$ is positive definite (semi-definite).

\section{Problem Formulation}\label{s2}
\subsection{System Model} \label{s2.2}
This paper considers a class of linear systems:
\begin{align}
  \label{equ:systemstate}
  \begin{split} x_{k+1} = \ & A x_{k} + B u_{k} + \omega_{k},  \\
    y_{k} = \ & C x_{k} + \nu_{k}, \ k \in \mathbb{N},
  \end{split}
\end{align}
where $x_{k} \in \mathbb{R}^{n}$, $u_k \in \mathbb{R}^{m}$, and $y_{k} \in \mathbb{R}^{p}$ denote the system state, input, and output at time step $k$, respectively; $A \in \mathbb{R}^{n \times n}$, $B \in \mathbb{R}^{n \times m}$, and $C \in \mathbb{R}^{p \times n}$ are unknown system state, input, and output matrices, respectively; $\omega_k \in \mathbb{R}^{n}  \sim  \mathcal{N}(0,\ Q)$ and $\nu_{k}  \in \mathbb{R}^{p} \sim  \mathcal{N}(0,\ R)$ are the system process and measurement noise with $Q \in \mathbb{R}^{n \times n}  > 0$ and $R \in \mathbb{R}^{p \times p} > 0$, respectively. The initial system state is denoted by $x_{0}  \sim  \mathcal{N}(\bar{x}_0,\ P_{0})$ with $\bar{x}_0 \in \mathbb{R}^{n}$ and $P_{0}  \in \mathbb{R}^{n \times n} > 0$. We assume that $x_{0}$, $\omega_{k}$, and $\nu_{k}$, $\forall k \in \mathbb{N}$, are mutually uncorrelated.
% and $A$ is non-singular. For most physical plants, the latter assumption is very mild because system \dref{equ:systemstate} can be regarded as the discretized form of a differential equation $\text{d} x = F_1 x + F_2 \text{d} w $ such that $A \triangleq  e^{F_1 h} $ is always invertible for any sufficiently small sampling period $h$ \cite{shi2014event}.

\begin{assumption} \label{assumptionstabilizable}
%System \eqref{equ:systemstate}, or equivalently,
$(C, \ A)$ is observable.
\end{assumption}

%
%We assume that $A$, $B$, and $C$ in \dref{equ:systemstate} are unknown, while the definition of the system state is given. This assumption is motivated by scenarios  where the system matrices are unknown but the state is clearly defined. If a system is described by an input-output relationship without a clearly defined state, the state estimation problem may be ill-posed, which does not fall within the research scope of this paper.

\subsection{Data Collection} \label{s2.11}

\begin{figure*}
\centering
   \subfigure[The pre-collected input-state-output trajectory.]{\includegraphics[scale=0.44]{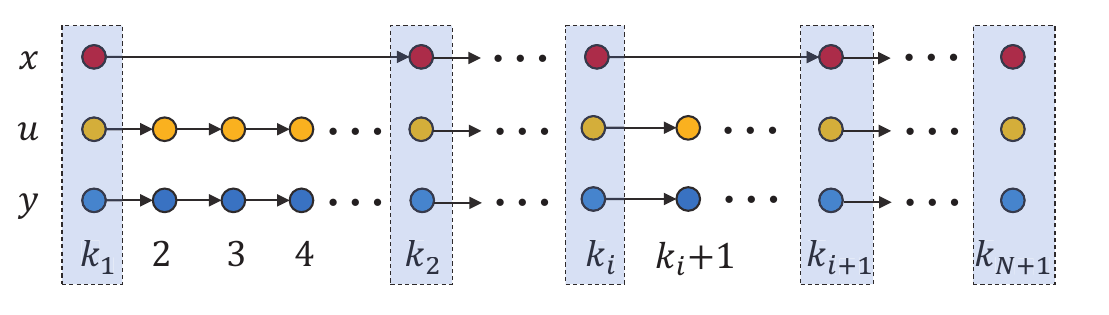}}
   \subfigure[The online input-state-output trajectory.]{\includegraphics[scale=0.47]{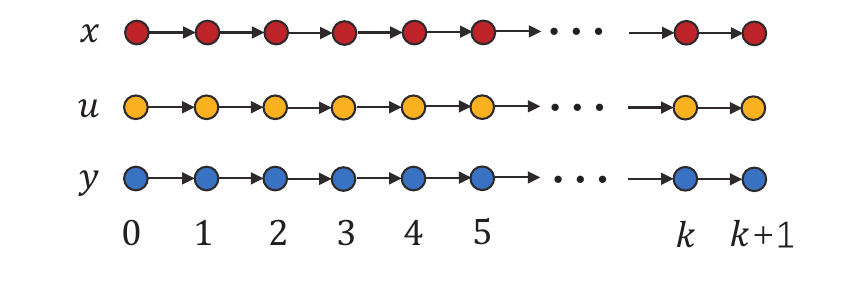}}
  \caption{The pre-collected and online system trajectories, where   red, yellow, and blue solid dots denote the state, input, and output, respectively. In Fig. (a), the states are sampled at time instants $k_1$, $k_2$, $\ldots$, $k_{N+1}$, and the inputs and outputs are sampled at every time instant from $k_1$ to $k_{N+1}$. This paper aims to estimate the online state at every time instant, using the pre-collected input-state-output trajectory and the online input-output trajectory.}
  \label{fig:sample}
\end{figure*}

%\begin{figure}[t]
%  \centering
%   {\includegraphics[scale=0.45]{sampling1}}
%  \caption{The pre-collected input-state-output sequence, where the red, yellow, and blue solid dots denotes the sampled state, input, and output, respectively. The state is sampled at time instant $k_i$, $i=1,\ldots,N+1$, and the input and output are sampled at time instant $k$, $k=1,\ldots,k_{N+1}$. }
%  \label{fig:sample}
%\end{figure}

{Suppose that we can pre-collect an input-state-output trajectory of  \dref{equ:systemstate}, as shown in Fig. \ref{fig:sample}(a), where the state sampling frequency is lower than the input and output sampling frequency. To be more specific, the sampled state sequence is denoted by
\begin{align} \label{equ:augmentinitial}
  x^{\text{p}} & =  [x_{k_1}^T, \ x_{k_2}^T, \ x_{k_3}^T, \ \ldots, \ x_{k_{N+1}}^T]^T,
  \end{align}
 where the superscript `p' denotes the pre-collected data.
Then, let $L_i = k_{i+1} - k_i $, $i \in \mathcal{V} \triangleq \{1$, $2$, $\ldots$, $N\}$. Without loss of generality, we assume that there exists a constant scalar $L$ such that $L_i \ge L$, $\forall i \in \mathcal{V}$. If $L_i < L$ for some  $i \in \mathcal{V}$, we can omit the sampled state $x_{k_{i+1}}$ and retain the next sampled state until the condition $L_i \ge L$ is satisfied. This assumption is practical, particularly when the state sampling frequency is much lower than the input-output sampling frequency. Next, we divided the input-output trajectory into $N$ segments and then extracted the first $L$ points from each segment, as
\begin{align} % \label{equ:eachdata}
u^{i,\text{p}}  & = [u_{k_i}^T, \ u_{k_i + 1}^T, \  \ldots, \ u_{k_i + L -1}^T]^T,  \notag \\
y^{i,\text{p}}  &  = [y_{k_i}^T, \ y_{k_i + 1}^T, \  \ldots, \ y_{k_i + L }^T]^T,
 \notag
\end{align}
where $i \in \mathcal{V} $ denotes the $i$-th segment. The augmented input and output data of all segments are denoted by
\begin{align} \label{equ:colletiondata}
  \begin{split}
  u^{\text{p}} & =  [(u^{1,\text{p}})^T, \ \ldots, \ (u^{N,\text{p}})^T]^T,  \\
  y^{\text{p}} & =  [(y^{1,\text{p}})^T, \ \ldots, \ (y^{N,\text{p}})^T]^T.
\end{split}
  \end{align}
Let $\omega^{\text{p}}$ and $\nu^{\text{p}}$ denote the corresponding process and measurement noise, respectively.}
%\begin{align}
%  \omega^{i,\text{p}} &= [\omega_{k_i}^T, \ \omega_{k_i + 1}^T, \  \ldots, \ \omega_{k_i + L -1}^T]^T,  \notag  \\
%  \nu^{i,\text{p}}  &=  [\nu_{k_i}^T, \ \nu_{k_i + 1}^T, \  \ldots, \ \nu_{k_i + L}^T]^T. \notag
%\end{align}
Further, we define the following notations
\begin{flalign} \label{pre-data}
  & U \triangleq [u^{1,\text{p}}, \ \ldots, \ u^{N,\text{p}}], & \ &  Y  \triangleq [y^{1,\text{p}}, \ \ldots, \ y^{N,\text{p}}], \notag \\
  & \Omega  \triangleq [\omega^{1,\text{p}}, \ \ldots, \ \omega^{N,\text{p}}], & \ &   V  \triangleq [\nu^{1,\text{p}}, \ \ldots, \ \nu^{N,p}], \\
  & X \triangleq  [x_{k_1}, \ \ldots, \ x_{k_{N}}].   \notag
\end{flalign}
%\begin{flalign}
%  & U = [u^{1,p}, \ \ldots, \ u^{N,p}],& \ &  Y  = [y^{1,p}, \ \ldots, \ y^{N,p}], \notag \\
%  & W  = [W^{1}, \ \ldots, \ W^{N}],& \ &    W^{i} = [(\omega_0^i)^T, \ \ldots, \ (\omega_{L-1}^i)^T]^T,\notag \\
%  & V  = [V^{1}, \ \ldots, \ V^{N}],& \ &   V^{i}  = [(\nu_0^i)^T, \ \ldots, \ (\nu_{L}^i)^T]^T, \notag \\
%  & {\color{green}X  = [\hat{x}_0^{1},\ \ldots, \ \hat{x}_0^{N}],}& \ &    \Xi  = [\xi ^{1}, \ \ldots, \ \xi^{N}]. \notag
%\end{flalign}
It follows from \dref{equ:systemstate} that
\begin{align} \label{equ:augdata}
  Y = G X  +  F U  + H \Omega + V,
\end{align}
where $F = H (I_{L} \otimes B)$,
\begin{align}
  G = &  \left [                %左括号
  \begin{array}{c}   %该矩阵一共3列，每一列都居中放置
    C  \\
    CA   \\
    CA^{2}    \\
    \vdots   \\
    CA^{L}    \\
  \end{array} \right ], \quad
  H =   \left [                %左括号
  \begin{array}{ccccc}   %该矩阵一共3列，每一列都居中放置
    0 & 0 & \cdots & 0  \\
    C & 0 &   & 0 \\
    CA & C &   & 0 \\
    \vdots  &  & \ddots  &    \vdots \\
    CA^{L-1} & CA^{L-2} & \cdots & C  \\
  \end{array} \right ]. \notag
\end{align}
%For clarity, let notations with superscript $i$ denote data collected from pre-collected trajectories (e.g., $y_h^i$), and others represent variables or measurements in the current filtering process (e.g., $y_k$).

\begin{assumption} \label{assumptionL}
    $L \ge   \max \{ n, m, p\} $.
\end{assumption}

\begin{remark}
The conditions $L \ge m$ and $L \ge p$ are assumed only for notational simplicity, while the condition $L \ge n$  is assumed to ensure the observability matrix has full column rank \cite[Theorem 6.DO1]{chen1984linear}. % When the observability index of system \dref{equ:systemstate}, denoted by $n_o$, is available, Assumption \ref{assumptionL} can be relaxed to $L \ge   n_o $.
\end{remark}

\begin{assumption} \label{assumptionx0u}
    $ \textup{rank}  \bigg  [
 \begin{array}{c}
   X \\
   U \\
 \end{array} \bigg ]   = n + Lm $.
\end{assumption}

\begin{remark}
To ensure that Assumption \ref{assumptionx0u} holds, we can first choose a sequence of $u^{\textup{p}} $ such that $ \textup{rank}  (U ) = Lm $, which can be realized by following the input design methods in \cite[Chapter 13.3]{ljung1998system}. Next, it suffices to ensure $ \textup{rank}  (X) = n $. According to \cite[Theorem 2.1]{coulson2021distributionally}, when the input $u^{\textup{p}} $ is exciting, the condition $ \textup{rank}  (X) = n $ holds.
\end{remark}

% In model-based estimation and control problems, it is always assumed that available. However, they are usually regarded as unknown information in the data-driven society, which should be identified from or equivalently interpreted by the pre-collected system trajectories. This paper focuses on the latter formulation.

{
\textbf{Motivating Example}: Continuous stirred-tank reactors (CSTRs) play a crucial role in industrial processes, particularly in chemical production and reaction engineering. State monitoring of CSTRs is essential for ensuring process efficiency, maintaining product quality, and preventing safety hazards. According to \cite{bequette2003process}, a CSTR for an exothermic reaction can be described by:
\begin{align}
& \dot{C}_{\textup{A}}   = \frac{q}{V}(C_{\textup{f}} - C_{\textup{A}} ) - k_0 \textup{exp} \Big ( \! - \frac{E}{RT} \Big) C_{\textup{A}}  , \notag \\
& \dot{T}  = \frac{q}{V}(T_{\textup{f}} - T) - \frac{k_0   \Delta H } {\rho C_\textup{p}} \textup{exp} \Big ( \! - \! \frac{E}{RT} \Big) C_{\textup{A}} + \frac{UA(T_\textup{c}-T)}{V\rho C_\textup{p}}, \notag
\end{align}
where $C_{\textup{A}}  \in \mathbb{R}$ is the reactant concentration; $T \in \mathbb{R}$ is the reactor temperature that can be measured; $T_\textup{c}  \in \mathbb{R}$ is the temperature of the coolant stream that can be manipulated; and see \cite{bequette2003process} for the definitions of other parameters. Note that the nominal operating setpoint of the CSTR corresponds to a steady state $C^{\textup{s}}$, $T^{\textup{s}} $, and $T_{\textup{c}}^{\textup{s}}$. By letting $x=[C_{\textup{A}}  -C^{\textup{s}}, \ T  -T^{\textup{s}} ]^T$, $u = T_\textup{c}$, and $y=T$ be the state, input, and output vectors, the dynamics of the CSTR near the nominal operating setpoint can be modeled as \dref{equ:systemstate} \cite{bequette2003process}, where $A \in \mathbb{R}^{2 \times 2} $ and $B \in \mathbb{R}^{2}$ are unknown system matrices. For this system, we can pre-collect an input-state-output trajectory. Specifically, we can measure temperature using sensors at a minute-level frequency, and measure reactant concentration through chemical analysis methods usually at an hourly frequency \cite{bequette2003process}. By doing so, we obtain a higher-frequency sampled input-output trajectory, denoted by $ u^{\text{p}}$ and $y^{\text{p}} $, and a lower-frequency sampled state trajectory, denoted by $x^{\text{p}}$. The objective is to estimate the online reactant concentration using the pre-collected system trajectory.}

This paper focuses on state estimation for a predefined state of an unknown system, as shown in the example above, requiring prior data related to this state. This differs from subspace identification or Willems' fundamental lemma-based control \cite{oymak2019non, coulson2019data,turan2021data,adachi2021dual}, which relies only on input-output data and does not require a specific state-space model. {We will also generalize the results to cases without prior state information, applicable to control tasks such as LQG control.}

\subsection{Problem Statement} \label{s2.3}
The objective of this paper is to estimate an online state trajectory of \dref{equ:systemstate} based on the pre-collected data $\{u^{\text{p}}$, $x^{\text{p}}$, $y^{\text{p}}\}$. The online trajectory is described by
 \begin{align} \label{equ:currentinput}
  \begin{split}
    u_{[0,k]}  & \triangleq [u_0^T, \  u_1^T, \ \ldots, \ u_{k}^T]^T, \\
    x_{[0,k+1]}   & \triangleq [x_0^T, \ x_1^T, \ \ldots, \ x_{k+1}^T]^T, \\
    y_{[1,k+1]}   & \triangleq [y_1^T, \ y_2^T, \ \ldots, \ y_{k+1}^T]^T, \  k \in \mathbb{N},
  \end{split}
 \end{align}
where $u_{[0,k]}$ and $y_{[1,k+1]}$ are known, as shown in Fig. \ref{fig:sample}(b).

{\bf Problem:} For system \dref{equ:systemstate} with unknown $A$, $B$, and $C$, design a filtering algorithm to estimate the state $ x_{[0,k+1]}$ defined in \dref{equ:currentinput}, using $\{u^{\text{p}}$, $x^{\text{p}}$, $y^{\text{p}}\}$ and $\{u_{[0,k]}$, $y_{[1,k+1]}\}$, defined in \dref{equ:augmentinitial},  \dref{equ:colletiondata},  and \dref{equ:currentinput}, respectively, as
 \begin{align} \label{equ:hatxaug}
\hat{x}_{[0,k+1]} = g_{k+1} ( u_{[0,k]}, y_{[1,k+1]}, u^{\text{p}}, y^{\text{p}}, \hat{x}_0^{\text{p}}), \ k \in \mathbb{N},
\end{align}
where
$\hat{x}_{[0,k+1]}  = [\hat{x}_0^T,  \ldots, \hat{x}_{k+1}^T]^T$
denotes the estimate of the online $x_{[0,k+1]}$ that corresponds to \dref{equ:systemstate}, and $g_{k+1}(\cdot)$ denotes the filtering algorithm. Moreover, the filtering performance should be quantitatively analyzed and compared with the Kalman filter based on known matrices $A$, $B$, and $C$.

\section{RDKF Design}  \label{s3}
In this section, we provide a framework for estimating the state in  \dref{equ:currentinput} using the pre-collected data $\{u^{\text{p}}$, $x^{\text{p}}$, $y^{\text{p}}\}$. Let ${\bf{x}}$, ${\bf{y}}$, and ${\bf{y^p}}$ be the random variables of   $\hat{x}_{[0,k+1]}$, $y_{[1,k+1]}$, and $y^\textup{p}$, respectively.
%Motivated by the fact that Kalman filter for system \dref{equ:systemstate} with known $A$, $B$ and $C$ can be derived using ML optimization \cite{rauch1965maximum},
Then, we define a joint probability density function of $\hat{x}_{[0,k+1]}$, $y_{[1,k+1]}$, and $y^\textup{p}$:
%the system states and outputs of the online observed trajectory \dref{equ:currentinput}, and the initial states and outputs of previous trajectories \dref{equ:colletiondata} and \dref{equ:augmentinitial}, as
\begin{align}
& f_{{\bf{x}},   {\bf{y}},   {\bf{y^p}}}  (\hat{x}_{[0,k+1]},  y_{[1,k+1]},   y^{\text{p}}) \notag \\
 \triangleq \ & f_{{\bf{x}},  {\bf{y}}, {\bf{y^p}}}  ({\bf{x}} \! = \! \hat{x}_{[0,k+1]}, {\bf{y}} \! = \! y_{[1,k+1]}, {\bf{y^p}} \! = \! y^{\text{p}}). \notag
\end{align}
% where the notations ${\bf{x}} = \hat{x}_{[0,k+1]}$, ${\bf{y}} = y_{[1,k+1]}$, and ${\bf{y^p}} =y^{\text{p}}$  mean that the random variables ${\bf{x}}$, ${\bf{y}}$, and ${\bf{y^\textup{p}}}$ take some deterministic values $\hat{x}_{[0,k+1]}$, $y_{[1,k+1]}$, and $y^{\text{p}}$, respectively.

\begin{lemma} \label{lemmaf}
 { For system \eqref{equ:systemstate}, the joint probability density function $f_{{\bf{x}},   {\bf{y}},   {\bf{y^p}}}  (\hat{x}_{[0,k+1]},  y_{[1,k+1]},   y^{\textup{p}})$ is equivalent to }
  \begin{align} % \label{equ:f}
    & \ f_{{\bf{x}},   {\bf{y}},   {\bf{y^p}}}  (\hat{x}_{[0,k+1]},  y_{[1,k+1]},   y^{\textup{p}})  = \textup{constant} \times  f_{\textup{off}} \times f_{\textup{on}}, \notag
   \end{align}
where
\begin{align} % \label{equ:f2}
 f_{\textup{off}} =     &  \prod_{i=1}^{N}  \prod_{h=0}^{L-1}  \textup{exp} \Big ( - \frac{1}{2} (\hat{\omega}_h^i)^T Q^{-1} \hat{\omega}_h^i  \Big )   \notag \\
  \times  &      \prod_{i=1}^{N}  \prod_{h=0}^{L} \textup{exp}  \Big ( - \frac{1}{2} (\hat{\nu}_{h}^i)^T R^{-1} \hat{\nu}_{h}^i \Big ),  \notag
   \end{align}
   and
\begin{align} % \label{equ:f3}
 f_{\textup{on}}
    = \ & \textup{exp} \Big ( - \frac{1}{2} (\hat{x}_0 - \bar{x}_0)^T P_{0}^{-1} (\hat{x}_0 - \bar{x}_0)   \Big ) \notag  \\
         \ \times & \prod_{t=0}^{k} \textup{exp} \Big ( - \frac{1}{2} \hat{\omega}_t^T Q^{-1} \hat{\omega}_t - \frac{1}{2} \hat{\nu}_{t+1}^T R^{-1} \hat{\nu}_{t+1}  \Big ) ,   \notag
   \end{align}
$\hat{\omega}_{[0,k]}  \triangleq  [\hat{\omega}_0^T$, $ \ldots$, $ \hat{\omega}_{k}^T]^T$ and $ \hat{\nu}_{[1,k+1]}  \triangleq   [\hat{\nu}_1^T$, $ \ldots$, $\hat{\nu}_{k+1}^T]^T$ are variables to approximate the system process and measurement noise, respectively, satisfying
\begin{align}
   \label{equ:constriantw}
    \begin{split}
       \hat{x}_{t+1} & =    A \hat{x}_{t} + B u_{t} + \hat{\omega}_{t},  \\
       y_{t} & =   C \hat{x}_{t} + \hat{\nu}_{t}, \ t =0, 1, \ldots, k+1,
  \end{split}
\end{align}
 and
  \begin{align}
   \hat{\omega}^{\textup{p}}   = & [(\hat{\omega}^1)^T, \ldots, (\hat{\omega}^N)^T]^T, \ \hat{\omega}^i   =  [( \hat{\omega}_0^i)^T, \ldots, ( \hat{\omega}_{L-1}^i)^T]^T, \notag \\
   \hat{\nu}^{\textup{p}}   =  & [(\hat{\nu}^1)^T, \ldots, (\hat{\nu}^N)^T]^T, \ \hat{\nu}^i   =  [( \hat{\nu}_0^i)^T, \ldots, ( \hat{\nu}_{L}^i)^T]^T, \notag
  \end{align}
  % with $\hat{\xi}^i$, $\hat{\omega}_h^i$  and $ \hat{\nu}_h^i$,  $\forall i \in \mathcal{V}$,
  are variables with their elements satisfying
  \begin{align} \label{equ:outputnoisydata1}
    y_{k_i + h} \! = \!  C A^h x_{k_i} \!+\! \sum_{l=1}^{h} C A^{l-1} (B u_{h-l}^i \!+ \!\hat{\omega}_{h-l}^i )   \!+ \! \hat{\nu}_h^i,
   \end{align}
 for all $h = 1, \ldots, L$,  and $ y_{k_i}=  C   x_{k_i}  + \hat{\nu}_0^i$.
\end{lemma}

%\begin{algorithm}[t]
%  \caption{ML-Based Data-Driven Filter.} \label{algorithm1}
%  \hspace*{0.02in}
%
%  {\bf Initialize:} collect $U$, $Y$, and $X$ defined in \dref{pre-data};
%
%  {\bf for $k =0, 1, \ldots$ do}
%\begin{algorithmic}[1]
%  \State solve \dref{equ:optimization1} to obtain the optimal values of variables $\hat{x}_{t}$, denoted by $\hat{x}_{t}^*$, $t = 0, \ldots, k$;
%
%  \State  let the state estimate of system \dref{equ:systemstate} at time step $k$ be $\hat{x}_{t}^*$ with $t = k$;
%\end{algorithmic}
%{\bf end for}
%
%\end{algorithm}

% , i.e.,
% \begin{align} % \label{equ:inf}
%  & \ \text{ln} f (\hat{x}_0, \hat{\omega}, \hat{\nu}, \hat{\xi}^{\text{p}}, \hat{\omega}^{\text{p}}, \hat{\nu}^{\text{p}} ) \notag \\
%   = &   -  \frac{1}{2} \sum_{t=0}^{k}  \hat{\omega}_t^T Q^{-1} \hat{\omega}_t  -  \frac{1}{2} \sum_{t=0}^{k}  \hat{\nu}_{t+1}^T R^{-1} \hat{\nu}_{t+1} + \text{constant} \notag \\
%  &  -  \frac{1}{2} \sum_{i=1}^{N}  \sum_{h=0}^{L-1}  (\hat{\omega}_h^i)^T Q^{-1} \hat{\omega}_h^i   -   \frac{1}{2} \sum_{i=1}^{N}  \sum_{h=0}^{L}   (\hat{\nu}_{h}^i)^T R^{-1} \hat{\nu}_{h}^i  \notag \\
%  &   - \sum_{i=1}^{N}  \frac{1}{2} (\hat{\xi}^i)^T P_{\xi}^{-1} \hat{\xi}^i   - \frac{1}{2} (\hat{x}_0 - \hat{x}_0)^T P_{0}^{-1} (\hat{x}_0 - \hat{x}_0) .  \notag
% \end{align}

The proof of Lemma \ref{lemmaf} is given in Appendix \ref{lemmafproof}.
According to \cite{rauch1965maximum}, when the system matrices $A$, $B$, and $C$ are  known, the {minimum mean-square error (MMSE)} state estimate for \dref{equ:systemstate} is the optimal solution to an ML optimization problem. In this paper, we consequently perform state estimation for \dref{equ:systemstate} by instead maximizing the likelihood function derived in Lemma \ref{lemmaf}, i.e.,
\noindent
\begin{align}  \tag{$\mathcal{P}^{\uppercase\expandafter{\romannumeral1}}_{k+1}$}
\begin{split}
 \min_{  \hat{x}_0,   \hat{\omega}^{\text{p}}, \hat{\nu}^{\text{p}}, \atop \hat{\omega}_{[0,k]}, \hat{\nu}_{[1,k+1]}}   &  -   \text{ln} f_{\textup{off}} -   \text{ln} f_{\textup{on}}
    \\ \text{s.t.}  \qquad   &    \quad \dref{equ:constriantw}   \ \text{and} \ \dref{equ:outputnoisydata1}.
  \end{split}
\end{align}
Let $\hat{x}_{0}^*$, $\hat{\omega}^{*}_{[0,k]}$, $\hat{\nu}^*_{[1,k+1]}$, $\hat{\omega}^{\text{p}*}$, and $\hat{\nu}^{\text{p}*}$ denote the solution to   $\mathcal{P}^{\uppercase\expandafter{\romannumeral1}}_{k+1}$. Let $\hat{x}_{[0,k+1]}^* = [(\hat{x}_0^*)^T$, $(\hat{x}_2^*)^T$, $\ldots$, $  (\hat{x}_{k+1}^*)^T]^T$ denote the state evolution of \dref{equ:constriantw} given these optimal variables.

\begin{algorithm}[t]
  \caption{RDKF} \label{algorithm2}
  \hspace*{0.02in}

{\bf Input:} $U$, $Y$, $X$, $u$, and $y$, defined in \dref{pre-data} and \dref{equ:currentinput};

{\bf Output:} $ \hat{x}_{k}$, $k \in \mathbb{N}$;

  \begin{algorithmic}[1]

  \State  compute matrices
  \begin{align}  \label{equ:ABCsharp}
        A_{\sharp}  & = G_{1,\sharp}^{\dagger} G_{2,\sharp}, \ B_{\sharp} = G_{1,\sharp}^{\dagger} G_{3,\sharp}, \ C_{\sharp} = G_{4,\sharp};
  \end{align}
  with
\begin{align} \label{equ:G1234}
\begin{split}
      G_{1,\sharp} &   =    Z_{\sharp} (1:Lp;1:n),    \\
      G_{2,\sharp} &   =    Z_{\sharp} (p+1:Lp+p;1:n),     \\
      G_{3,\sharp} &   =    Z_{\sharp}(p+1:Lp+p;n+1:2n),     \\
      G_{4,\sharp} & = G_{1,\sharp}(1\!:\!p;1\!:\!n), \ Z_{\sharp} =  Y  ( [  X^T,   U^T]^T) ^{\dagger};
\end{split}
    \end{align}

  \State  {\bf for $k =0, 1, \ldots$ do}
  {
    \begin{align}
    & \hat{x}_{k+1|k} =  \hat{A}_k \hat{x}_{k} +  \hat{B}_k u_k,  \notag \\
    & \hat{x}_{k+1} = \hat{x}_{k+1|k} + L_{k+1} (y_{k+1} - \hat{C}_k \hat{x}_{k+1|k} ), \notag \\
    & \label{equ:subml} L_{k+1} = \bar{P}_{k+1}  \hat{C}_k^T (\hat{R} + \hat{C}_k  \bar{P}_{k+1} \hat{C}_k^T)^{-1},   \\
    & \bar{P}_{k+1} =  A_{\sharp} \hat{P}_k A_{\sharp}^T + Q,  \notag \\
    & P_{k+1} =  \bar{P}_{k+1} \! - \! \bar{P}_{k+1} \hat{C}_k^T (\hat{R} + \hat{C}_k
    \bar{P}_{k+1} \hat{C}_k^T)^{-1} \hat{C}_k \bar{P}_{k+1}, \notag
  \end{align}
where the  filter parameters are defined as
  \begin{align}
    & \hat{A}_k \!=\!  A_{\sharp} (I - \lambda \psi_A^2 \epsilon \hat{P}_k ), \ \hat{B}_k \!=\!  B_{\sharp}  - \lambda  \psi_B^2 \epsilon \hat{P}_k, \ \hat{C}_k \!=\!  C_{\sharp},    \notag \\
    &  \hat{R} \!=\!  R  - \lambda^{-1}   \epsilon C_{\sharp} C_{\sharp}^T,  \   \hat{P}_k \!=\! ((P^{\sharp}_{k})^{-1} + \lambda  \psi_A^2 \epsilon)^{-1}, \notag
  \end{align}
 $ \hat{x}_{k+1}$ is the estimate of $x_{k+1}$ in \dref{equ:currentinput}, $\hat{x}_0 = \bar{x}_0$, $\lambda$ is any scalar greater than $\| \epsilon   C_{\sharp}^T R^{-1} C_{\sharp}\|_2$; and $\psi_A$, $\psi_B$, and $\epsilon$ are given in \dref{equ:epsi}.}  \\
 {\bf end for}
\end{algorithmic}
\end{algorithm}

Note that the optimization problem $\mathcal{P}^{\uppercase\expandafter{\romannumeral1}}_{k+1}$ is nonlinear and non-convex, and cannot be solved in general. We slightly modify $\mathcal{P}^{\uppercase\expandafter{\romannumeral1}}_{k+1}$ to get a problem solvable with a recursive algorithm and with a solution close to the optimal one with known $A$, $B$, and $C$. Specifically, we modify $\mathcal{P}^{\uppercase\expandafter{\romannumeral1}}_{k+1}$ as
\begin{align}  \tag{$\mathcal{P}^{\uppercase\expandafter{\romannumeral2}}_{k+1}$}
\begin{split}
 \min_{\hat{x}_0, \hat{\omega}_{[0,k]},  \atop  \hat{\nu}_{[1,k+1]}} \ \max_{ \hat{\omega}^{\text{p}} \in B_{\epsilon_1} (\hat{\omega}^{\text{p*}}),  \atop  \hat{\nu}^{\text{p}} \in B_{\epsilon_2} (\hat{\nu}^{\text{p*}}) }  &    - \text{ln} f_{\textup{on}}
    \\ \text{s.t.}  \qquad   &    \quad \dref{equ:constriantw} ,
  \end{split}
\end{align}
with $B_{\epsilon_1} (\hat{\omega}^{\text{p*}}) = \{\hat{\omega}^{\text{p}} | \| \hat{\omega}^{\text{p}}-\hat{\omega}^{\text{p*}}\|_2 \leq \epsilon_1 \}  $ and $B_{\epsilon_2} (\hat{\nu}^{\text{p*}}) = \{\hat{\nu}^{\text{p}} | \| \hat{\nu}^{\text{p}}-\hat{\nu}^{\text{p*}}\|_2 \leq \epsilon_2 \}  $, where $\hat{\omega}^{\text{p}*}$ and $\hat{\nu}^{\text{p}*}$ are solution to
\begin{align}  \tag{$\mathcal{P}^{\uppercase\expandafter{\romannumeral2}}_{0}$}
\begin{split}
 \min_{ \hat{\omega}^{\text{p}}, \hat{\nu}^{\text{p}} }   &  -   \text{ln} f_{\textup{off}}, \quad     \text{s.t.}  \    \dref{equ:outputnoisydata1}, \notag
  \end{split}
\end{align}
and $\epsilon_1$ and $\epsilon_2$ are the upper bounds of $\| {\omega}^{\text{p}}-\hat{\omega}^{\text{p*}}\|_2$ and $ \| {\nu}^{\text{p}}-\hat{\nu}^{\text{p*}}\|_2$, respectively.

To solve $\mathcal{P}^{\uppercase\expandafter{\romannumeral2}}_{k+1}$, we propose the RDKF algorithm. {Let us explain the main stages of the algorithm. First, when regarding $A$, $B$, and $C$ as unknown variables, the optimal solution to $\mathcal{P}^{\uppercase\expandafter{\romannumeral2}}_{0}$ can be directly derived as $\hat{\omega}^{\text{p}}=0$ and $\hat{\nu}^{\text{p}}=0$. Then, if Assumption \ref{assumptionx0u} holds, substituting the solution into \dref{equ:outputnoisydata1} gives
 \begin{align} \label{equ:YGHsharp}
 Z_{\sharp}   \triangleq  [G_{\sharp}    \quad   F_{\sharp}  ]   =    Y  ( [  X^T,   U^T]^T) ^{\dagger},
\end{align}
where $Y$, $X$, and $U$ are given in \dref{equ:augdata}, and $G_{\sharp}$ and $F_{\sharp}$ are the estimates of $G$ and $F$, respectively. Let $A_{\sharp}$, $B_{\sharp}$, and $C_{\sharp}$ be the estimates of  $A$, $B$, and $C$, respectively. We assume that $G_{\sharp}$ and $F_{\sharp}$ have the same structures with respect to $A_{\sharp}$, $B_{\sharp}$ and $C_{\sharp}$ as $G$ and $F$ have with respect to $A$, $B$ and $C$, respectively. Hence, the following equations
\begin{align} %
 G_{1,\sharp} A_{\sharp}  \! = \!   G_{2,\sharp},   G_{1,\sharp} B_{\sharp}  \! =\!   G_{3,\sharp},   C_{\sharp} \! = \! G_{1,\sharp}(1:m;1:n), \notag
\end{align}
hold. If $ G_{1,\sharp}$ has full column rank, we obtain \dref{equ:ABCsharp}. When there exist positive scalars $\psi_A$, $\psi_B$, and $\epsilon$ such that $\|A - A_{\sharp}\|_2 \leq \psi_A \epsilon$, $\|B - B_{\sharp}\|_2 \leq \psi_B \epsilon$, and $\|C - C_{\sharp}\|_2 \leq \epsilon$ (the values of $\psi_A$, $\psi_B$, and $\epsilon$ will be derived in \dref{equ:epsi}), $\mathcal{P}^{\uppercase\expandafter{\romannumeral2}}_{k+1}$ can be slightly modified as
\begin{align*}
  &  \min_{ \hat{\omega}_{[0,k]}, \hat{\nu}_{[1,k+1]} }   \  \max_{ \|\Delta_A\|_2 \leq \psi_A \epsilon,  \atop \|\Delta_B\|_2 \leq \psi_B \epsilon,   \|\Delta_C\|_2 \leq \epsilon  }  - \text{ln} f_{\textup{on}}  \\
    \text{s.t.}  \quad      &   \hat{x}_{t+1}    =    ( A_{\sharp} + \Delta_A ) \hat{x}_{t} + (B_{\sharp}  + \Delta_B ) u_{t} + \hat{\omega}_{t},  \\
    & y_{t}   =  ( C_{\sharp} + \Delta_C ) \hat{x}_{t} + \hat{\nu}_{t}, \ t =0, 1, \ldots, k+1.
\end{align*}
By utilizing the regularized least-squares method in \cite{935054}, an explicit solution to the above optimization problem is derived as \dref{equ:subml} in Algorithm \ref{algorithm2}.}

\begin{remark}
Maximum likelihood estimation is a fundamental principle for state estimation, e.g., in switching \cite{alessandri2010maximum}, time-varying \cite{poncela2013automatic}, and nonlinear systems \cite{marelli2015asymptotic}. This paper employs this principle to deal with state estimation of unknown systems, where the new likelihood function derived in Lemma \ref{lemmaf} incorporates both the pre-collected data and the online measurements.
\end{remark}

{
When no prior state can be collected, the RDKF can be directly extended to estimate the state of a balanced realization, applicable to control tasks like LQG control. To move on, we restack $U$ and $Y$ in \dref{pre-data} by replacing $L$ with $2L$, and modify Assumption \ref{assumptionx0u} as
\begin{assumption} \label{assumptionu}
    $ \textup{rank} ( U )  =  2 Lm $.
\end{assumption}
Similarly to system identification \cite{tsiamis2020sample,zheng2020non}, we assume the initial state of pre-collected input-output \mbox{trajectory \dref{equ:colletiondata}} to be either zero or unknown zero-mean random noise. In this case, the RDKF proposed in Algorithm \ref{algorithm2} remains valid, except that $Z_{\sharp}$ in \dref{equ:G1234} is replaced by
 \begin{align} \label{equ:Znew}
 Z_{\sharp}^{\textup{new}} & \! = \! [ G_{\sharp}^{\textup{new}}     \   F_{\sharp}^{\textup{new}}(1\!:\!Lp\!+\!p;\!1\!:\!Lm) ], \ F_{\sharp}^{\textup{new}} \!=  \!Y U^{\dagger}, \notag \\
 G_{\sharp}^{\textup{new}} & \! =\! U^\textup{s}_1  \Sigma^{1/2}_1, \ {F}_{\sharp,H}^{\textup{new}} \! \xlongequal [] {\textup{SVD}} \! [U^\textup{s}_1 \ U^\textup{s}_2] \! \Big [ \!
        \begin{array}{cc}
          {\Sigma_1} & {} \\ \!
          {} & { \Sigma_2} \\
        \end{array} \!\Big ]  \Big [\!
        \begin{array}{c}
          V^\textup{s}_1 \\
          V^\textup{s}_2 \\
        \end{array} \!\Big ],
\end{align}
where $ {F}_{\sharp,H}^{\textup{new}} $ is the Hankel form of   ${F}_{\sharp}^{\textup{new}} $ \cite{tsiamis2019finite,oymak2019non};  ``SVD" denotes the singular value decomposition; $\Sigma_1 \in \mathbb{R}^{n \times n}$ contains the $n$-largest singular values. Let us explain the derivation process of the new result as follows. Similarly to cases with prior states, we still solve  $\mathcal{P}^{\uppercase\expandafter{\romannumeral2}}_{k}$ to design the RDKF for cases without prior states. In the new case, \dref{equ:YGHsharp} reduces to $ F_{\sharp}^{\textup{new}} =  Y U^{\dagger}$, where $F_{\sharp}^{\textup{new}}$ is the estimate of $F$. Note that $F$ is a block-Toeplitz matrix, which can be reshaped as a Hankel matrix $F_{H}$ satisfying $F_{H} = G G_{B} $, where $G$ is the observability matrix and $G_{B}$ is the controllability matrix \cite[Section 3]{tsiamis2019finite}. Hence, we can reshape $F_{\sharp}^{\textup{new}}$ to get a Hankel matrix ${F}_{\sharp,H}^{\textup{new}}$. By taking the SVD  of ${F}_{\sharp,H}^{\textup{new}}$, the controllability matrix of a balanced realization can be derived as $ G_{\sharp}^{\textup{new}}$ in \dref{equ:Znew}. As a result, $Z_{\sharp}$ in \dref{equ:G1234} is replaced by  $Z_{\sharp}^{\textup{new}}$ in \dref{equ:Znew}, while the other steps in Algorithm \ref{algorithm2} remain valid.
}

\section{Data Informativity Analysis} \label{s4}
This section is aimed at analyzing the informativity of the pre-collected data required for performing the RDKF. Similarly to \dref{equ:YGHsharp}, let $Z  \triangleq  [G     \    F]$, and $e_Z = Z_{\sharp} - Z$ be the error between $Z$ and $Z_{\sharp}$. It follows from \dref{equ:augdata} that
\begin{align} \label{equ:ez}
  e_Z  =  ( H \Omega + V)  \left [
        \begin{array}{c}
          X \\
          U \\
        \end{array} \right ]^T \bigg ( \! \left [
        \begin{array}{c}
          X \\
          U \\
        \end{array} \right ] \left [
          \begin{array}{c}
            X \\
            U \\
          \end{array} \right ]^T \! \bigg )^{-1},
\end{align}
which is obtained by substituting the explicit expression of $ ([X^T, U^T]^T )^{\dagger}$. Similarly to \cite{tsiamis2020sample}, we assume that the system is stable and $L_i \gg L$. In the experiments of generating data, let $u_{h}^{i} \sim  \mathcal{N}(0,\ S)$ with $S \in \mathbb{R}^{m \times m} >0$. In this case, $x_{k_i}$ is a Gaussian variable, and we denote $x_{k_i} \sim  \mathcal{N}(0,\ P_{x})$ with $P_{x} \in \mathbb{R}^{n \times n} > 0$. When $L_i \gg L$ and the system state matrix $A$ is stable, it is reasonable to assume that $u_{h}^{i}$ and $x_{k_i}$ are mutually uncorrelated, $\forall h = k_i, k_i + 1, \ldots, k_i+L-1$, $\forall i \in \mathcal{V} $. Let $Q = \sigma_{\omega}^2 I_n$, $R = \sigma_{\nu}^2 I_p$, $P_{\xi} = \sigma_{\xi}^2 I_n$,  $S = \sigma_{u}^2 I_m$, and $P_{x} = \sigma_{x}^2 I_n$ for notational simplicity.

\begin{proposition}\label{thm3}
 Consider system \eqref{equ:systemstate} with the collected data $U$, $Y$, and $X$ defined in \eqref{pre-data}. Suppose \mbox{Assumption \ref{assumptionx0u}} holds. If $N \ge 32L^2 \textup{log}(27/\delta)$, then $
    \| e_Z \|_2 \leq \mathcal{O}   ( \sqrt{ {1}/{N }}  )$
 holds with probability at least $1-\delta$ with $\delta \in (0, \ 1)$.
\end{proposition}

The proof of Proposition \ref{thm3} is given in \mbox{Appendix \ref{thm3proof}}. Let $G_{1}$, $G_{2}$, and $G_{3}$ represent the corresponding block matrices in $Z$ as $G_{1,\sharp}$, $G_{2,\sharp}$, and $G_{3,\sharp}$ in $Z_{\sharp}$, respectively, where $G_{1,\sharp}$, $G_{2,\sharp}$, and $G_{3,\sharp}$ are defined in Algorithm \ref{algorithm2}. Subsequently, we have the following corollary.

\begin{corollary} \label{cor2}
 Consider system \eqref{equ:systemstate} with the pre-collected data $U$, $Y$, and $X$,  defined in \eqref{pre-data}. Suppose \mbox{Assumption \ref{assumptionx0u}} holds. For any scalars $\epsilon \in(0$, $1)$ and $\delta \in(0$, $1)$, there always exists a positive integer $N_0(\epsilon,\delta)$ such that if $N \ge N_0 (\epsilon,\delta)$, then
 $$ \| G_{1} - G_{1,\sharp} \|_2   \leq  \epsilon, \
    \| G_{2} - G_{2,\sharp} \|_2  \leq  \epsilon,  \
    \| G_{3} - G_{3,\sharp} \|_2   \leq  \epsilon,  $$
hold with probability at least $1-\delta$.
\end{corollary}

%Till now, the errors between $\{G_{1}$, $G_{2}$, $G_{3}\}$ and $\{G_{1,\sharp}$, $G_{2,\sharp}$, $G_{3,\sharp}\}$ have been analyzed quantitatively. Further,

{
From the proof of Proposition \ref{thm3}, one feasible $N_0(\epsilon,\delta)$ is given by
\begin{align}  \label{equ:NG}
   N_0 (\epsilon,\delta)\!  = \!  32 L^2  \textup{log}(27/\delta) \!  \times \! \max   \{ \! 1,    2 {M_Z^2}/{\epsilon^2} \! +\!  2 \alpha_0^2  \! \},
\end{align}
where $M_Z$ is a positive constant defined as
\begin{align} \label{Mz}
  M_Z =  \alpha_0  \| G_{\sharp} \|_2     + \beta_0,
\end{align}
with $\alpha_0 = \sigma_{\text{max}}   \sigma_{\omega}  L/\sigma_{\text{min}}^2$, $\beta_0 = \sigma_{\text{max}}  \sigma_{\nu}/ \sigma_{\text{min}}^2$, $\sigma_{\text{max}} = \max \{ \sigma_{x}, \sigma_{u} \}$, and $\sigma_{\text{min}} = \min \{ \sigma_{x}, \sigma_{u} \}$.}
The above result reveals an explicit relation between the estimation error bound $\epsilon$ and the number of the pre-collected states $N$.
%It shows that larger noise covariances in the pre-collected data will result in a larger $\epsilon$ for a given $N$.
Based on Corollary \ref{cor2}, we can determine the number $N$ required for ensuring the full column rank of $G_{1,\sharp}$  for performing the proposed RDKF.

\begin{theorem}\label{thm4}
Suppose Assumptions \ref{assumptionstabilizable}, \ref{assumptionL} and  \ref{assumptionx0u} hold. If $N \ge N_0 (\epsilon,\delta)$ and $ \epsilon < \epsilon_0 $, where $N_0 (\epsilon,\delta)$ is defined in \dref{equ:NG} and $\epsilon_0$ is defined as
\begin{align} \label{equ:epsilon0}
  \epsilon_0 \triangleq \sqrt{\|G_{1}\|_2^2 + \lambda_{\min}(G_{1}^T G_{1}) } - \|G_{1}\|_2,
\end{align}
then $G_{1,\sharp}$ has full column rank with probability at least $1-\delta$.
\end{theorem}

 Theorem \ref{thm4} is proved in Appendix \ref{thm4proof}. Theorem \ref{thm4} reveals that $G_{1,\sharp}$ has full column rank with any high probability if conditions in the Theorem \ref{thm4} are met. % Particularly, if system \dref{equ:systemstate} is noise-free, i.e., $\omega_k = \nu_k =0$, $\forall k \in \mathbb{N}$, $\epsilon_0 $ and $N_0 (\epsilon,\delta)$ can be chosen as $0$ and $n + Lm$, respectively. In this case, if Assumptions \ref{assumptionstabilizable}, \ref{assumptionL} and  \ref{assumptionx0u} hold, then  $G_{1,\sharp}$ consistently has full column rank and $e_Z$ in Proposition \ref{thm3} is always $0$.

% number of the pre-collected system trajectories is large enough. % By virtue of this fact, we can collect sufficient data to meet the requirement on $G_{1,\sharp}$ for the designed RDKF \dref{equ:subml}.

\section{RDKF Performance Evaluation} \label{s5}

This section presents the performance gap between the proposed RDKF and the optimal Kalman filter based on known system matrices.

\begin{theorem}\label{thm5}
   Suppose that Assumptions \ref{assumptionstabilizable}, \ref{assumptionL}, and  \ref{assumptionx0u} hold.  When $N \ge N_0 (\epsilon,\delta)$ and $\epsilon < \epsilon_0$  with $N_0(\epsilon,\delta)$ and $\epsilon_0$ being defined in \dref{equ:NG} and \dref{equ:epsilon0}, respectively, then {
   $$     \| A -  A_{\sharp} \|_2   \leq \psi_A \epsilon, \
    \| B - B_{\sharp}\|_2    \leq \psi_B \epsilon, \
    \| C -  C_{\sharp} \|_2    \leq \epsilon,$$}
simultaneously hold with probability at least $1-\delta$, where $\psi_A$ and $\psi_B$ are positive constants defined in \dref{equ:epsi}.
\end{theorem}

 Theorem \ref{thm5} is proved in Appendix \ref{thm5proof}. Theorem \ref{thm5} offers the upper bounds for the estimation errors of system matrices. {Considering a given $N \ge L_0 \triangleq  32 L^2  \textup{log}(27/\delta)  $, it follows from \dref{equ:NG} and the proof of Theorem \ref{thm5} that the values of $\epsilon$, $\psi_A$, and $\psi_B$ is given by
\begin{align} \label{equ:epsi}
\begin{split}
\epsilon & =  \sqrt{ {2M_Z^2 L_0 }/({N -  2 \alpha_0^2 L_0})}, \\
\psi_A & = \| G_{1,\sharp}^{\dagger} \|_2    (  \| A_{\sharp} \|_2   + 1)/(1 - \| G_{1,\sharp}^{\dagger} \epsilon \|_2 ),     \\
\psi_B & = \| G_{1,\sharp}^{\dagger} \|_2    (  \| B_{\sharp} \|_2   + 1)/(1 - \| G_{1,\sharp}^{\dagger} \epsilon \|_2 ),
\end{split}
\end{align}
which are used for the RDKF presented in Algorithm \ref{algorithm2}.}

According to Theorems \ref{thm4} and \ref{thm5}, $G_{1,\sharp}$ has full column rank with probability at least $1-\delta$ when $N \ge N_0 (\epsilon,\delta)$ and $ \epsilon < \epsilon_0 $. Hence, $(C$, $ A_{\sharp})$ is observable \cite[Theorem 6.DO1]{chen1984linear}. Further, $\hat{P}_{k}$, $\bar{P}_{k}$, and ${P}_{k}$ in \dref{equ:subml} exponentially converge to the unique solution to  \cite{935054}
\begin{align}   \label{equ:subKalmanhatP1}
  \begin{split}
   \hat{P}_{\sharp} = & \ ( P_{\sharp}^{-1} + \lambda  \psi_A^2 \epsilon)^{-1}, \ \bar{P}_{\sharp} =  A_{\sharp} \hat{P}_{\sharp}  A_{\sharp}^T + Q,     \\
  P_{\sharp}  =  & \ \bar{P}_{\sharp}  - \bar{P}_{\sharp}  C_{\sharp}^T (\hat{R} + C_{\sharp} \bar{P}_{\sharp}  C_{\sharp}^T)^{-1} C_{\sharp} \bar{P}_{\sharp}, \\
\end{split}
\end{align}
with probability at least $1-\delta$ when $N \ge N_0 (\epsilon,\delta)$ and $ \epsilon < \epsilon_0 $. We assume that the filter parameters in \dref{equ:subml} are in their steady states for notational simplicity. In addition, the steady-state Kalman filter for \dref{equ:systemstate} with known $A$, $B$, and $C$ is given by \cite{anderson2005optimal}
 \begin{align}
    \hat{x}_{k+1|k} = & \  A \hat{x}_{k} + B u_k, \notag  \\
   \hat{x}_{k+1} = & \ \hat{x}_{k+1|k} + L  (y_{k+1} - C \hat{x}_{k+1|k} ), \notag   \\
    \label{equ:Kalman}  L =  & \  \bar{P} C^T (R + C  \bar{P}  C^T)^{-1},   \\
     \bar{P}  =  & \ A P A^T + Q,    \notag \\
     P  =  & \ \bar{P}  - \bar{P}  C^T (R + C \bar{P}  C^T)^{-1} C \bar{P}, \notag
 \end{align}
 where $\hat{x}_{k+1|k}$ and $\hat{x}_{k+1}$ are the  {\it a priori} and {\it a posteriori} MMSE estimates, respectively; and $L $, $\bar{P} $, and $P $ are filter parameters. Further, an assumption on the state $x_k$ is needed, which holds for many systems with closed-loop controllers.
 %Note that $\bar{P}_{k+1}$ defined converges to $\bar{P} $ exponentially, where
% % \begin{align}
%    \begin{split} \label{equ:steadybarP}
%    \bar{P}  =  & \ A P A^T + Q,   \\
%     P  =  & \ \bar{P}  - \bar{P}  C^T (R + C \bar{P}  C^T)^{-1} C \bar{P}.
%  \end{split}
%  \end{align}
%In the following, the error between $\bar{P}^{\sharp}_{k+1}$ and $\bar P$ and the error between  ${P}^{\sharp}_{k+1}$ and $P$ are quantified.
%
%\begin{proposition}\label{thm6}
% Suppose Assumptions \ref{assumptionstabilizable}, \ref{assumptionL} and  \ref{assumptionx0u} hold. If $N \ge N_0 (\epsilon,\delta)$ and $\epsilon < \epsilon_0$ with $N_0(\epsilon,\delta)$ and $\epsilon_0$ being defined in \dref{equ:NG} and \dref{equ:epsilon0}, respectively, then
%  \begin{align}
%    \| \bar{P}_{\sharp} - \bar{P}  \|_2 & \leq   \mathcal{O} \bigg ( \sqrt{\frac{\log(1/\delta)}{N }} \bigg), \notag
%  \end{align}
%  or
%  \begin{align}
%      \|  {P}_{\sharp} -  {P}  \|_2 & \leq   \mathcal{O} \bigg ( \sqrt{\frac{\log(1/\delta)}{N }} \bigg), \notag
%  \end{align}
%holds with probability at least $1-\delta$.
%\end{proposition}
%The proof of Proposition \ref{thm6} is provided in Appendix \ref{thm6proof}. Proposition \ref{thm6} will contribute to the performance evaluation of the proposed RDKF in the next subsection.

\begin{assumption}  \label{xbound}
There exists a positive definite matrix $\Pi$ such that $\mathbb{E}\{x_{k}x_{k}^T\} \leq \Pi$, $\forall k \in \mathbb{N}$.
\end{assumption}

%Assumption \ref{xbound} holds for many physical systems with closed-loop controllers. In particular, two types of data-driven controllers will be designed for system \dref{equ:systemstate} to guarantee this assumption in Section \ref{s6}.
%

Let $e_{k} = \hat{x}_{k} - x_{k} $ be the estimation error of the RDKF at step $k$, and $P_{e,k}\triangleq \mathbb{E}\{e_{k}e_{k}^T\}$. A result about the filtering performance of the proposed RDKF is given as follows.

\begin{theorem}\label{thm7}
  Consider system \eqref{equ:systemstate} with the pre-collected data $U$, $Y$, and $X$, defined in \eqref{pre-data}. Suppose Assumptions \ref{assumptionstabilizable}, \ref{assumptionL},  \ref{assumptionx0u}, and \ref{xbound} hold. When $N \ge N_0 (\epsilon,\delta)$ and $\epsilon < \epsilon_0$  with $N_0(\epsilon,\delta)$ and $\epsilon_0$ being defined in \dref{equ:NG} and \dref{equ:epsilon0}, respectively, then
  \begin{align}
    \| P_{e,\infty} - P  \|_2 & \leq  \mathcal{O} \big ( \sqrt{ {1}/{N}} \big),  \notag
  \end{align}
holds with probability at least $1 - \delta$.
\end{theorem}

 The proof of Theorem \ref{thm7} is provided in Appendix \ref{thm7proof}. Theorem \ref{thm7} provides a sample-complexity bound for the performance gap between the designed RDKF \dref{equ:subml} and the traditional Kalman filter \dref{equ:Kalman} with known $A$, $B$, and $C$. Particularly, we have $ \| P_{e,\infty} - P  \|_2 \rightarrow 0$ when $N \rightarrow \infty$. In this case, the performance of the designed RDKF is as good as the model-based Kalman filter.

 {If no prior state is available, as derived in Section \ref{s3}, we modify Algorithm \ref{algorithm2} by replacing $Z_{\sharp}$ in \dref{equ:G1234} with $Z_{\sharp}^{\textup{new}}$ in \dref{equ:Znew}. In this case, the state estimate corresponds to a balanced realization, which can be used for control tasks like LQG control. Similarly to Theorem \ref{thm5}, we can also derive sample-complexity upper bounds for the learning errors of $A, B, C$. By denoting $P$ in \dref{equ:Kalman} for this case as $P_{\textup{b}}$, we have the following corollary.

\begin{corollary} \label{corlast}
  Consider system \eqref{equ:systemstate} with the pre-collected data $U$ and $Y$ defined in \eqref{pre-data}. Suppose Assumptions \ref{assumptionstabilizable}, \ref{assumptionL},  \ref{assumptionu}, and \ref{xbound} hold. In addition, $Z_{\sharp}$ in Algorithm \ref{algorithm2} is replaced with $Z_{\sharp}^{\textup{new}}$ in \dref{equ:Znew}. For any scalars $\epsilon \in(0$, $1)$ and $\delta \in(0$, $1)$, there always exists a positive integer $N_0(\epsilon,\delta)$ such that if $N \ge N_0 (\epsilon,\delta)$, then
  \begin{align}
    \| P_{e,\infty} - P_{\textup{b}} \|_2 & \leq  \mathcal{O} \big ( \sqrt{ {1}/{N}} \big),  \notag
  \end{align}
holds with probability at least $1 - \delta$.
\end{corollary}

\begin{remark}
Corollary \ref{corlast} reveals that the designed RDKF can be adapted for cases where prior states are unavailable, while ensuring a sample complexity bound for the filtering performance. In particular, we introduce a novel approach that uses the estimation error covariance as the metric for the sample complexity analysis. This metric offers an intuitive assessment of the performance gap between the proposed RDKF and the optimal Kalman filter based on known system matrices.
\end{remark}
}

\section{Simulation}\label{s7}

%\begin{figure}[t]
%  \centering
%   \subfigure[$C_{\textup{A}}$ and $\hat{C}_{\textup{A}}$.]{\includegraphics[scale=0.45]{concent}}
%   \subfigure[$T$ and $\hat{T}$.]{\includegraphics[scale=0.45]{tempe}}
%  \caption{{The real states and the estimates using the proposed RDKF, where $\hat{C}_{\textup{A}}$ and $\hat{T}$ are the estimates of $C_{\textup{A}}$ and $T$, respectively.}}
%  \label{fig:motoronce}
%\end{figure}

{

In this section, the effectiveness of the proposed RDKF is illustrated using a simulation of a CSTR \cite{bequette2003process}. The dynamics of the CSTR is described by a linear state-space model, see Section \ref{s2.11}. Similarly to \cite{bequette2003process}, consider that the nominal operating setpoint of the CSTR corresponds to a steady state $C^{\textup{s}} = 0.5  \ \textup{mol/l}$, $T^{\textup{s}} = 350 \ \textup{K}$, and $T_{\textup{c}}^{\textup{s}} =  300 \ \textup{K}$. Let $x=[C_{\textup{A}} -C^{\textup{s}}, \ T  -T^{\textup{s}} ]^T$, $u = T_\textup{c}$, and $y=T$ be the state, input, and output vectors. By utilizing a sampling time of $t_s = 0.1 \ \textup{min}$, the dynamics of the CSTR is modeled as \dref{equ:systemstate} \cite{sui2010linear} with
\begin{align}
A \! = \! \left[ \! { \! \begin{array}{cc}
   {0.7776} & { -0.0045} \\
   {26.6186} & { 1.8555}  \end{array} \! } \! \right], \ B \! = \!  \left[ \! { \! \begin{array}{c}
    { -0.0004} \\
    {0.2907}  \end{array} \! } \! \right], \ C \! = \!    [0 \quad 1]. \notag
\end{align}
Let $\sigma_{\omega} = 0.01$ and $\sigma_{\nu} = 0.1$. Assume that the above matrices $A$ and $B$ are unknown.} Suppose that we can collect an input-state-output sequence $u^{\text{p}}$, $x^{\text{p}}$, and $y^{\text{p}}$ defined in \dref{equ:colletiondata} and \dref{equ:augmentinitial}, where the parameters are chosen as $N = 100$, $L = 5$, $\sigma_{x} = 0.4$, and $\sigma_{u} = 2$. Based on these data, we apply the proposed RDKF to estimate an online trajectory of the CSTR with $x_0=[0.4, \ 5]^T$ and $u_k=-8y_k$. To proceed, two types of errors are defined:
\begin{align}
  \textup{MSE}(h) & \! = \! \frac{1}{50} \sum_{k = 1}^{50} \|x_k^h - \hat{x}_k^h\|_2^2, \  \textup{AMSE}   \! = \! \frac{1}{N_{t}} \sum_{h = 1}^{N_{t}} \textup{MSE}(h) ,  \notag
\end{align}
where the notation $h$ denotes the $h$-th trial, and $N_{t} = 200$ denotes the number of  Monte Carlo trials. %; $\textup{MSE}(h)$ refers to the average mean-square estimation error of the \mbox{$h$-th} trial; and  $\textup{AMSE}$ refers to the average mean-square estimation error of all trials.

\begin{figure}[t]
  \centering
   {\includegraphics[scale=0.4]{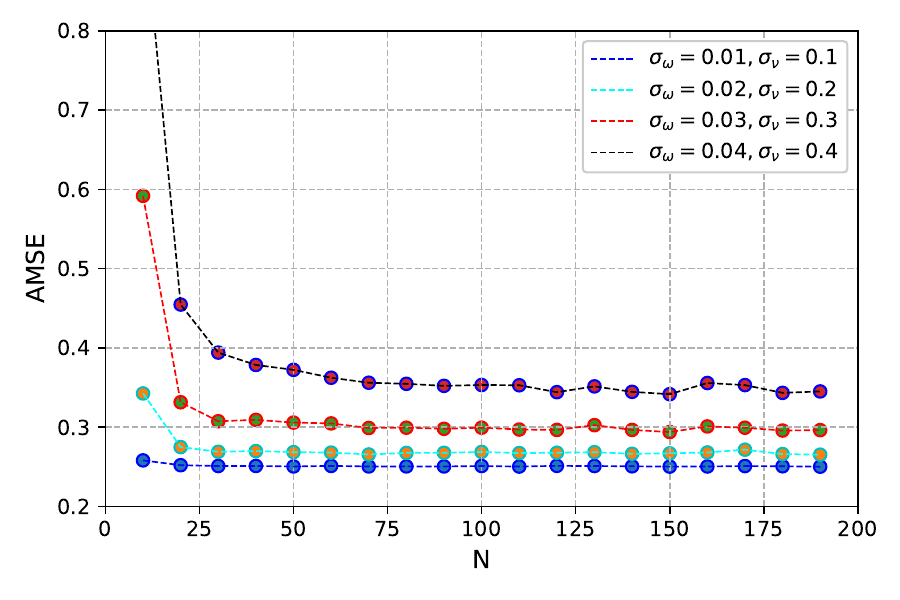}}
  \caption{{The values of $\textup{AMSE}$ under different numbers of samples and magnitudes of noise using the proposed RDKF, where the number $N$ is set as $10, 20, \ldots, 190$, respectively.} }
  \label{fig:comparison}
\end{figure}

\begin{figure}[t]
  \centering
  {\includegraphics[scale=0.35]{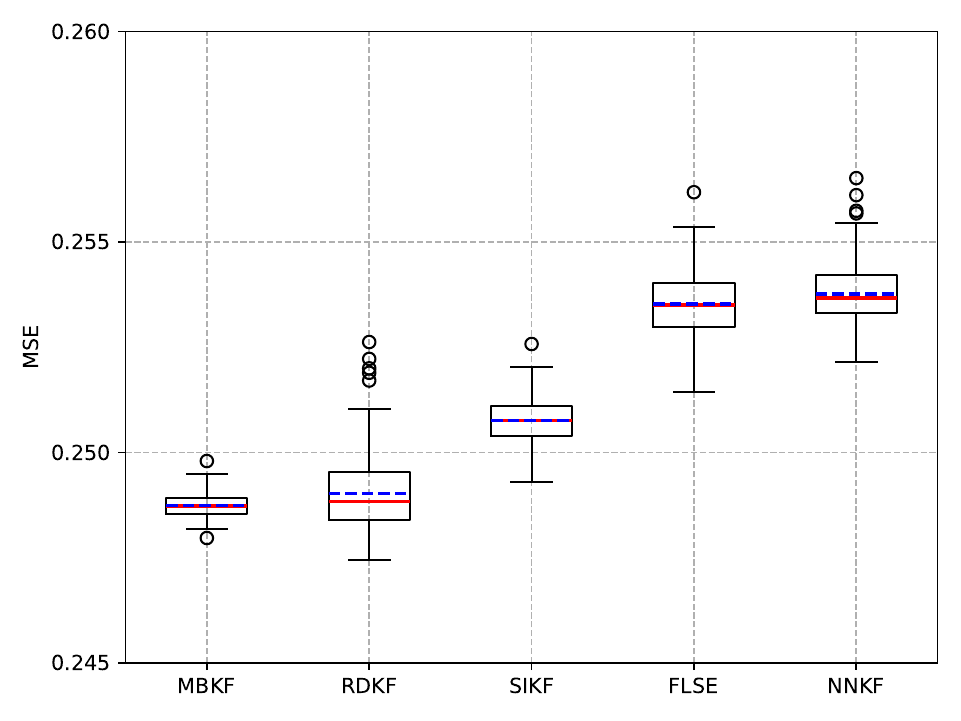}}
  \caption{{The values of MSE of 200 Monte Carlo trials using different filtering methods, where the red solid line and the blue dotted line denote the median and mean, respectively; the tops and bottoms of each box represent the 25th and 75th percentiles, respectively; and black circles denote outliers beyond 1.5 times the interquartile range. }}
  \label{fig:com}
\end{figure}

{The filtering performance of the CSTR by the designed RDKF is illustrated in \mbox{Figs. \ref{fig:comparison} and \ref{fig:com}.}
%\mbox{Fig. \ref{fig:motoronce}} shows that the estimate differs from the true state at the initial step but rapidly converges to the true one.
Fig. \ref{fig:comparison} illustrates that the filtering performance of the proposed RDKF is improved with an increasing number of samples, which coincides with Theorem \ref{thm7}. Moreover, it also shows that the smaller noise magnitudes leads to a better filtering performance. In addition, four relevant filtering methods in the literature are included for comparison, namely the model-based Kalman filter using known system matrices (MBKF) \cite{anderson2005optimal}, system identification-based Kalman filter (SIKF) \cite{verhaegen2007filtering}, Willems' fundamental lemma-based state estimator (FLSE) \cite{liu2023learning}, and NNs-based Kalman filter (NNKF) \cite{Guy2022}. It is worth mentioning that the SIKF, FLSE, and NNKF use a prior state trajectory sampled at every time step, whereas the RDKF employs a lower-frequency sampled state trajectory. Even so, it is shown in \mbox{Fig. \ref{fig:com}} that the RDKF demonstrates better filtering performance and superior capability in handling noisy data. The above observations illustrate the theoretical results obtained in this paper.
}

\section{Conclusion}\label{s8}
This paper proposed a new RDKF for a class of unknown linear systems, where a prior input-output trajectory sampled at a higher frequency and a prior state trajectory sampled at a lower frequency are available. Specifically, an ML optimization problem has been formulated and solved to construct the state estimate. A sample-complexity upper bound has been derived for the performance gap between the designed RDKF and the Kalman filter with known system parameters. Simulations have demonstrated the effectiveness of the theoretical results.

\section{Appendix}

\subsection{Proof of Lemma \ref{lemmaf}} \label{lemmafproof}
First,  the probability density functions of other variables are also simplified similarly to  $f_{{\bf{x}}, {\bf{y}}, {\bf{y^p}}}  (\hat{x}_{[0,k+1]}$, $y_{[1,k+1]}$, $ y^\text{p}) $, e.g.,  $f_{  {\bf{y^p}}} (   {\bf{y^p}} =   y^\text{p} ) = f_{  {\bf{y^p}}} (  y^\text{p}  ) $. Since $\hat{x}_{[0,k+1]}$ and $y_{[1,k+1]}$ are independent on the pre-collected data $u^{\text{p}}$ and $y^\text{p}$, we have
\begin{align} \label{equ:conditionp}
  & f_{{\bf{x}},  {\bf{y}},  {\bf{y^p}}}  (\hat{x}_{[0,k+1]}, y_{[1,k+1]},   y^\text{p})  \notag \\
  =  & \  f_{{\bf{x}},   {\bf{y}}}  (\hat{x}_{[0,k+1]}, y_{[1,k+1]}| u^{\text{p}},  y^{\text{p}})  f_{  {\bf{y^p}}} ( y^{\text{p}}|u^{\text{p}}),
\end{align}
where $f_{{\bf{x}},   {\bf{y}}}  (\hat{x}_{[0,k+1]}, y_{[1,k+1]}| u^{\text{p}},  y^{\text{p}})  $ is the conditional probability density function of ${\bf{x}} = \hat{x}_{[0,k+1]}$ and  ${\bf{y}} = y_{[1,k+1]}$ given  ${\bf{u^p}} = u^{\text{p}}$ and ${\bf{y^p}} = y^{\text{p}}$.  In the following, the explicit expressions of $f_{{\bf{x}},  {\bf{y}}}  (\hat{x}_{[0,k+1]}, y_{[1,k+1]}|  u^{\text{p}}, y^{\text{p}})$ and $ f_{  {\bf{y^p}}} ( y^{\text{p}}|u^{\text{p}})$ are deduced.
By utilizing Bayes' rule and the Markovian structure in \dref{equ:systemstate}, $ f_{{\bf{x}},  {\bf{y}}}  (\hat{x}_{[0,k+1]}, y_{[1,k+1]}|  u^{\text{p}}, y^{\text{p}}) $  can be derived that
\begin{align} % \label{equ:conditioneach}
  & f_{{\bf{x}},  {\bf{y}}}  (\hat{x}_{[0,k+1]}, y_{[1,k+1]}|  u^{\text{p}}, y^{\text{p}})  \notag \\
  =  \ & f_{\bf{x_0}} ( \hat{x}_{0}) \prod_{t=0}^{k} f_{{\bf{x_{t+1}}}, {\bf{y_{t+1}}} } ( \hat{x}_{t+1},  y_{t+1}| \hat{x}_{t}, u, u^{\text{p}}, y^{\text{p}}),   \notag
\end{align}
where $\hat{x}_{t+1}$ and $y_{t+1}$ are vectors satisfying \dref{equ:constriantw}.
Substituting \dref{equ:constriantw} into the above equation yields
\begin{align} \label{equ:conditionwv}
  & f_{{\bf{x}},  {\bf{y}}}  (\hat{x}_{[0,k+1]}, y_{[1,k+1]}|  u^{\text{p}}, y^{\text{p}}) \notag \\
   = & \ f_{\bf{x_0}} ( \hat{x}_{0}) \prod_{t=0}^{k} f_{\textbf{${\omega}_t$}} ( \hat{{\omega}}_t ) f_{ {\nu}_{t+1}} ( \hat{\nu}_{t+1} ),
\end{align}
where $\hat{{\omega}}_t$ and $\hat{\nu}_{t+1}$ are vectors defined in \dref{equ:constriantw}. Moreover, according to the definitions of the initial state and system noise in Section \ref{s2.2},  their probability density functions can be explicitly expressed by
\begin{align} % \label{equ:noisedistribution}
  f_{\bf{s}}(s) = \frac{1}{2 \pi^{\text{dim}/2} |\Sigma|^{1/2}} \text{exp} \bigg ( - \frac{1}{2} (s - \mu)^T \Sigma^{-1} (s - \mu)   \bigg ), \notag
\end{align}
where $s$ represents elements in $\{\hat{x}_0$, $\hat{\omega}_t$, $\hat{\nu}_{t+1}\}$  with $\mu$ and $\Sigma$ corresponding to $\{\bar{x}_0$, $0$, $0\}$ and $\{P_0$, $Q$, $R\}$, respectively; and `$\text{dim}$' denotes the dimension of $s$. Hence, it follows from \dref{equ:conditionwv} and the above equation that
\begin{align} \label{equ:f1distribution}
  & f_{{\bf{x}},  {\bf{y}}}  (\hat{x}_{[0,k+1]}, y_{[1,k+1]}|  u^{\text{p}}, y^{\text{p}})  = \text{constant} \times   f_{\textup{on}} .
\end{align}
For the second term $  f_{  {\bf{y^p}}} ( y^{\text{p}}|u^{\text{p}})$ in \dref{equ:conditionp}, we can derive from  \dref{equ:colletiondata} that
$ f_{  {\bf{y^p}}} ( y^{\text{p}}|u^{\text{p}}) =  \prod_{i=1}^{N}   f_{ {\bf{y^{i,\text{p}}}}} ( y^{i,\text{p}}|u^{i,\text{p}})$.
% Since the exact value of $u^{i,\text{p}}$ is pre-set, $f_{{\bf{u^{\text{p}}}}, {\bf{y^p}}} (u^{\text{p}}, y^p) $ can further be written in the following conditional form
% with
% \begin{align}
%   f_{{\bf{u^{i,\text{p}}}}, {\bf{y^{i,\text{p}}}}} (u^{i,\text{p}}, y^{i,\text{p}}) =  f_{{\bf{u^{i,\text{p}}}}, {\bf{y^{i,\text{p}}}}} (y^{i,\text{p}}|u^{i,\text{p}})  . \notag
% \end{align}
Further, according to \dref{equ:outputnoisydata1}, when $x_{k_i}$ and $u^{i,\text{p}} $ are given, we have $
  f_{ {\bf{y^{i,\text{p}}}}} ( y^{i,\text{p}}|u^{i,\text{p}}) =
  \prod_{h=0}^{L-1}  f_{{\bf{{\omega}_h^i }}} (\hat{\omega}_h^i ) \prod_{h=0}^{L}  f_{{\bf{{\nu}_h^i }}} (\hat{\nu}_h^i ),  $
where  $\hat{\omega}_h^i$ and $ \hat{\nu}_h^i$ satisfy \dref{equ:outputnoisydata1}. Similarly to the derivation process in \dref{equ:f1distribution}, the expression of $ f_{  {\bf{y^p}}} ( y^{\text{p}}|u^{\text{p}}) $ can be computed as $
  f_{  {\bf{y^p}}} ( y^{\text{p}}|u^{\text{p}})  = \text{constant} \times   f_{\textup{off}}$.
Now, combining \dref{equ:f1distribution} with the above equation gives rise to Lemma \ref{lemmaf}.

% \subsection{Proof of Theorem \ref{thm1}} \label{thm1proof}
%  First, when the optimal solutions of ${\xi}^{\text{p}}$, ${\omega}^{\text{p}}$ and ${\nu}^{\text{p}}$ to \dref{equ:optimization1} are given, the problem \dref{equ:optimization1} can be rewritten as
%  \begin{align}
%   &   \minimize_{  \hat{x}_0, \hat{\omega}, \hat{\nu}, A_*, B_*, C_*}  \qquad - \text{ln} f (\hat{x}_0, \hat{\omega}, \hat{\nu},A_*, B_*, C_*)  \notag   \\
%   &   \text{s.t.}  \quad   \hat{x}_{t+1}  =   A_* \hat{x}_t + B_* u_t + \hat{\omega}_t,  \notag \\
%      & \qquad y_{t}  =      C_* \hat{x}_{t} + \hat{\nu}_{t},   \ t = 0,  1,  \ldots,  k+1, \notag \\
%      & \quad Y = G_* X +  H_* (I_{L} \otimes B_*) U  + G_* \hat{\Xi}_* + H_* \hat{W}_* + \hat{V}_*. \notag
% \end{align}
% Given any $A_*$, $B_*$ and $C_*$ satisfying the last constraint, the above problem can be re-formulated as
% \begin{align}
%   \minimize_{  \hat{x}_0, \hat{\omega}, \hat{\nu}}   &  \qquad  - \text{ln} f (\hat{x}_0, \hat{\omega}, \hat{\nu}|A_*, B_*, C_*)  \notag   \\
%     \text{s.t.}  \qquad  &   \hat{x}_{t+1}  =   A_* \hat{x}_t + B_* u_t + \hat{\omega}_t,  \notag \\
%      & y_{t}  =      C_* \hat{x}_{t} + \hat{\nu}_{t},   \  t = 0,  1,  \ldots,  k+1, \notag
% \end{align}
% which is the same as the traditional optimization problem in Section \ref{s2.2}. Subsequently, the optimal state estimate $\hat{x}$ has the form of \dref{equ:optimalml}.

\subsection{Proof of Proposition \ref{thm3}} \label{thm3proof}

First, according to \dref{equ:ez}, it can be directly derived that
\begin{align} \label{equ:ezinequality}
    \|e_Z\|_2 \leq  & \    (     \|  H \Omega  [  X^T,   U^T] \|_2  +    \|  V [  X^T,   U^T]  \|_2     )  \notag \\
        &  \times     \|   (  [  X^T,   U^T]^T   [  X^T,   U^T]   )^{-1}    \|_2.
\end{align}
In the following, the three terms on the right side of \dref{equ:ezinequality} are analyzed. For the first term,  with probability at least $1-\delta/3$,
\begin{align}
  & \    \|  H \Omega   [  X^T,   U^T]     \|_2
   \leq    \| H \|_2   \|  \Omega   [  X^T,   U^T]  \|_2 \notag \\
    \leq &  \ 4 \| H \|_2 \sigma_{\omega} \sigma_{\text{max}} \sqrt{N(n +Ln + Lm)\text{log}(27/\delta)}, \notag
\end{align}
holds when $N \ge 2 (n +Ln + Lm) \text{log}(4/\delta)$, where $\sigma_{\text{max}} = \max \{ \sigma_{x}$, $\sigma_{u} \}$ and the second ``$\leq$'' is derived using \mbox{Lemma 1} in \cite{zheng2020non,dean2020sample}. Similarly, the second term
  \begin{align}
    &  \big \|  V    [  X^T,   U^T]  \big  \|_2  \notag \\
  \leq & \ 4  \sigma_{\nu} \sigma_{\text{max}} \sqrt{N(n + p +Lp + Lm)\text{log}(27/\delta)}, \notag
  \end{align}
holds with probability at least $1-\delta/3$ when $N \ge 2 (n + p +Lp + Lm) \text{log}(3/\delta)$. By applying Lemma 2 in \cite{zheng2020non,dean2020sample}, the last term
\begin{align}
 & \ \sqrt{\lambda_{\text{min}}    (   [  X^T,   U^T]^T   [  X^T,   U^T]     )} \notag \\
 \ge  & \ \sigma_{\text{min}}  \Big ( \sqrt{N} - \sqrt{n+Lm} - \sqrt{2 \text{log}(3/\delta)} \Big )
 \ge     \frac{1}{2} \sigma_{\text{min}} \sqrt{N}, \notag
\end{align}
holds with probability at least $1-\delta/3$, where $\sigma_{\text{min}} = \min \{ \sigma_{x}$, $\sigma_{u} \}$, and the second ``$\ge$'' holds when $N \ge 8(n+Lm) + 16 \text{log}(3/\delta) $. Hence, the last term  on the right side of \dref{equ:ezinequality} can be relaxed as
\begin{align}
     \|   (  [  X^T,   U^T]^T   [  X^T,   U^T]   )^{-1}    \|_2
 \leq & \   {4}/({N\sigma_{\text{min}}^{2}}) . \notag
\end{align}
Using the union bound, it can be derived that
 \begin{align} % \label{equ:ezbound}
  \| e_Z \|_2 < M_Z \sqrt{ {\text{log}(27/\delta)}/{N }}, \notag
 \end{align}
holds with probability at least $1-\delta$ when $N \ge 8(n+Lm) + 2(Lp + Lm + Ln + p + n + 3) \text{log}(3/\delta)$, where $ M_Z $ is a constant defined below \dref{equ:NG}, and Assumption \ref{assumptionL} is used such that $2n +Lm \leq L (2 + m)$,  $n +Ln + Lm < L (1+n+m)$ and $n + p +Lp + Lm < L (1+2p+m)$. Noting that we have $\text{log}(10/\delta) \ge 1$, the condition $N \ge 8(n+Lm) + 2(Lp + Lm + Ln + p + n + 3) \text{log}(3/\delta)$ can be further relaxed as $N \ge 32L^2 \text{log}(10/\delta)$. Now, the proof of Proposition \ref{thm3} is complete.

\subsection{Proof of Theorem \ref{thm4}} \label{thm4proof}
  First of all, note that
   \begin{align}
   & \  G_1^T G_1 -  G_{1,\sharp}^T G_{1,\sharp}  \notag \\
 % =  & \  G_1^T G_1  - G_{1,\sharp}^T G_{1} + G_{1,\sharp}^T G_{1} -  G_{1,\sharp}^T G_{1,\sharp}    \notag \\
%  =  & \  ( G_1 - G_{1,\sharp})^T G_1  + G_{1,\sharp}^T ( G_1 - G_{1,\sharp})  \notag \\
  =  & \  ( G_1 - G_{1,\sharp})^T G_1 + [(G_{1,\sharp} - G_{1}) + G_{1} ]^T ( G_1 - G_{1,\sharp})   \notag \\
  \leq  & \ \| G_1 - G_{1,\sharp}  \|_2 \|G_1 \|_2 I_n +  ( \|G_{1,\sharp} - G_{1}\|_2    + \|G_{1}\|_2 )\notag \\
  & \  \times \|G_{1,\sharp} - G_{1}\|_2 I_n  .  \notag
   \end{align}
  Then, it follows from Corollary \ref{cor2} that, with probability at least $1-\delta$, $  G_1^T G_1 - G_{1,\sharp}^T G_{1,\sharp}
  \leq  \epsilon^2 I_n + 2 \epsilon \|G_{1}\|_2 I_n$ holds,
  when $N \ge N_0 (\epsilon,\delta)$. Further, we have
  \begin{align}
     \lambda_{\text{min}}(G_{1}^T G_{1}) I_n \leq &  G_{1}^T G_{1} =   G_{1}^T G_{1} - G_{1,\sharp}^T G_{1,\sharp} + G_{1,\sharp}^T G_{1,\sharp}   \notag \\
   \leq &    \epsilon^2 I_n + 2 \epsilon \|G_{1}\|_2 I_n + G_{1,\sharp}^T G_{1,\sharp}. \notag
    \end{align}
  Besides, according to \cite[Theorem 6.DO1]{chen1984linear}, $G_1^T G_1> 0$ when Assumptions \ref{assumptionstabilizable} and \ref{assumptionL} hold, which indicates $\lambda_{\text{min}}(G_{1}^T G_{1}) > 0$   holds. All together, with probability at least $1-\delta$, $ G_{1,\sharp}^T G_{1,\sharp}  \ge  \Big [ \lambda_{\text{min}}(G_{1}^T G_{1})  -  \epsilon^2  -  2 \epsilon \|G_{1}\|_2 \Big ] I_n >0$ holds, when $N \ge N_0 (\epsilon,\delta)$ and $
  \epsilon < \sqrt{\|G_{1}\|_2^2 + \lambda_{\text{min}}(G_{1}^T G_{1}) } - \|G_{1}\|_2$. This ensures that $ G_{1,\sharp}$ has full column rank.

\subsection{Proof of Theorem \ref{thm5}} \label{thm5proof}
 It follows from \dref{equ:augdata} that
 \begin{align}
    A - A_{\sharp}   %& \ G_1^{\dagger} G_{2}  -  G_{1,\sharp}^{\dagger} G_{2,\sharp}  \notag \\
%    = & \ G_1^{\dagger} G_{2} - G_{1,\sharp}^{\dagger} G_{2} + G_{1,\sharp}^{\dagger} G_{2} -  G_{1,\sharp}^{\dagger} G_{2,\sharp}  \notag \\
    = & \ (G_1^{\dagger}   - G_{1,\sharp}^{\dagger} ) G_{1} A + G_{1,\sharp}^{\dagger} (G_{2} -    G_{2,\sharp} )  \notag \\
    = & \ G_{1,\sharp}^{\dagger} (G_{1,\sharp}   - G_{1} )   A + G_{1,\sharp}^{\dagger} (G_{2} -    G_{2,\sharp} ),  \notag
 \end{align}
 where the last ``$=$" is based on $G_{1,\sharp}^{\dagger} G_{1,\sharp} = G_{1}^{\dagger} G_{1} = I_n $.
 According to the results revealed in Corollary \ref{cor2},
 $  \| A - A_{\sharp} \|_2  \leq    \| G_{1,\sharp}^{\dagger} \|_2    (  \| A \|_2   +1   ) \epsilon$ holds when $N \ge N_0 (\epsilon,\delta)$ with $  \epsilon < \epsilon_0$ in \dref{equ:NG}. Since $G_{1,\sharp}^{\dagger} = (G_{1,\sharp}^T G_{1,\sharp})^{-1} G_{1,\sharp}^T$, we have
 \begin{align}
   \| G_{1,\sharp}^{\dagger} \|_2  = & \ \sqrt{ \lambda_{\text{max}} \big ((G_{1,\sharp}^{\dagger})^T G_{1,\sharp}^{\dagger} \big )}
   =   \sqrt{ \lambda_{\text{max}} \big ( G_{1,\sharp}^{\dagger} (G_{1,\sharp}^{\dagger})^T \big )}     \notag  \\
   = & \ \sqrt{ \lambda_{\text{max}} \big (  (G_{1,\sharp}^T G_{1,\sharp})^{-1}  \big )}     =   \sqrt{ 1/\lambda_{\text{min}} \big (   G_{1,\sharp}^T G_{1,\sharp} \big )}     \notag \\
   \leq & \ \sqrt{  {1}/({   \lambda_{\text{min}}(G_{1}^T G_{1})  -  \epsilon^2  -  2 \epsilon \|G_{1}\|_2 }}). \notag
 \end{align}
 Hence, $\| A \|_2   \| G_{1,\sharp}^{\dagger} \|_2 + \| G_{1,\sharp}^{\dagger} \|_2$ is uniformly bounded.  Similarly, with probability at least $1-\delta$,
 \begin{align}
   & \ \| B - B_{\sharp} \|_2
 \leq    \| G_{1,\sharp}^{\dagger} \|_2    (  \| B \|_2   +1   ) \epsilon, \notag
 \end{align}
 and $ \| C - G_{4,\sharp} \|_2  \leq   \| G_{1} - G_{1,\sharp} \|_2   \leq  \epsilon$ hold. Hence, Theorem \ref{thm5} is proved.

\subsection{Proof of Theorem \ref{thm7}} \label{thm7proof}
When the filter parameters of the designed RDKF are in steady states, it follows from Theorem \ref{thm5} and \dref{equ:subKalmanhatP1} that
\begin{align}
& \| \hat{A} - A \|_2, \| \hat{B} - B \|_2, \| \hat{C} - C \|_2,   \| \hat{R} - R \|_2, \|  {P}_{\sharp} -  \hat{P}_{\sharp}  \|_2, \notag
  \end{align}
are upper bounded by $\mathcal{O}   ( \sqrt{ {1}/{N }}  )$
 with probability at least $1-\delta$, when $N \ge N_0 (\epsilon,\delta)$ and $\epsilon < \epsilon_0$, since $\epsilon \sim \mathcal{O}   ( \sqrt{ {1}/{N }}  )$.

First, we consider Case 1): $ \| \hat{B} - B \|_2 \!=\! \| \hat{C} - C \|_2 \!=\!  \| \hat{R} - R \|_2 \!=\! \| {P}_{\sharp} -  \hat{P}_{\sharp}  \|_2 \!=\!  0$. In this case, the designed RDKF can be rewritten as
 \begin{align}
    \hat{x}_{k+1|k} = & \ \hat{A} \hat{x}_{k} + B u_k, \notag  \\
   \hat{x}_{k+1} = & \ \hat{x}_{k+1|k} + L^{\sharp}  (y_{k+1} - C \hat{x}_{k+1|k} ), \notag   \\
    \label{equ:Kalmannewone}  L^{\sharp} =  & \  \bar{P}_{\sharp} C^T (R + C  \bar{P}_{\sharp}  C^T)^{-1},   \\
     \bar{P}_{\sharp}  =  & \ A_{\sharp} P_{\sharp} A_{\sharp}^T + Q,    \notag \\
     P_{\sharp}  =  & \ \bar{P}_{\sharp}  - \bar{P}_{\sharp}  C^T (R + C \bar{P}_{\sharp}  C^T)^{-1} C \bar{P}_{\sharp}. \notag
 \end{align}
Note that $P_{\sharp}$ and $P$ in \dref{equ:Kalmannewone} and \dref{equ:Kalman} can be rewritten as
   \begin{align}
   P^{-1}  \! =  \! ( \bar{P}^{-1}  +  C^T  R^{-1}  C   )^{-1} , \
         P_{\sharp}^{-1}  \! =  \!   ( \bar{P}_{\sharp}^{-1}   +  C^T  R^{-1}  C   )^{-1}.  \notag
    \end{align}
    Then, it follows from \dref{equ:Kalmannewone} and \dref{equ:Kalman} that
   \begin{align} \label{equ:errorbarP}
    & \bar{P}_{\sharp} -  \bar{P}  \\
    =  &  A_{\sharp} ( \bar{P}^{-1}_{\sharp}  +  C^T  R^{-1}  C   )^{-1} A_{\sharp} - A ( \bar{P}^{-1}  +  C^T  R^{-1}  C   )^{-1} A  \notag \\
    =  & A_{\sharp} ( \bar{P}^{-1}_{\sharp}  +  C^T  R^{-1}  C   )^{-1} A_{\sharp} - A_{\sharp} ( \bar{P}^{-1}_{\sharp}  +  C^T  R^{-1}  C   )^{-1} A  \notag \\
       + & A_{\sharp} ( \bar{P}^{-1}   +  C^T  R^{-1}  C   )^{-1} A  - A ( \bar{P}^{-1}   +  C^T  R^{-1}  C   )^{-1} A  \notag \\
     + &  A_{\sharp} ( \bar{P}^{-1}_{\sharp}  +  C^T  R^{-1}  C   )^{-1} A - A_{\sharp} ( \bar{P}^{-1}  +  C^T  R^{-1}  C   )^{-1} A.  \notag
    \end{align}
   The first four terms on the right side of the second ``='' in \dref{equ:errorbarP} satisfy
   \begin{align}
       & A_{\sharp} ( \bar{P}^{-1}_{\sharp}  +  C^T  R^{-1}  C   )^{-1} A_{\sharp} - A_{\sharp} ( \bar{P}^{-1}_{\sharp}  +  C^T  R^{-1}  C   )^{-1} A  \notag \\
       & +  A_{\sharp} ( \bar{P}^{-1}   +  C^T  R^{-1}  C   )^{-1} A  - A ( \bar{P}^{-1}   +  C^T  R^{-1}  C   )^{-1} A  \notag \\
     =   &  A_{\sharp}  ( \bar{P}^{-1}_{\sharp}  +  C^T  R^{-1}  C   )^{-1} (A_{\sharp} - A)  \notag \\
     & \qquad  +   (A_{\sharp}  - A ) ( \bar{P}^{-1}_{\sharp}  +  C^T  R^{-1}  C   )^{-1}   A    \notag \\
     \leq   &  \|  A_{\sharp} ( \bar{P}^{-1}_{\sharp}  +  C^T  R^{-1}  C   )^{-1}\|_2  \| A_{\sharp}  - A   \|_2 I_n      \notag \\
     &  \qquad  + \| ( \bar{P}^{-1}_{\sharp}  +  C^T  R^{-1}  C   )^{-1} A \|_2   \| A_{\sharp}  - A   \|_2 I_n \notag \\
     \leq   &  M_{1} \epsilon I_n, \notag
     \end{align}
  with probability at least $1-\delta$, where $M_1$ is a constant, when $N \ge N_0 (\epsilon,\delta)$ and $\epsilon < \epsilon_0$. For the last two terms in \dref{equ:errorbarP}, we have
     \begin{align}
        & A_{\sharp} \Big ( ( \bar{P}^{-1}_{\sharp}  +  C^T  R^{-1}  C   )^{-1}  -  ( \bar{P}^{-1}  +  C^T  R^{-1}  C   )^{-1} \Big ) A   \notag  \\
       = & A_{\sharp}  ( \bar{P}^{-1}_{\sharp}  +  C^T  R^{-1}  C   )^{-1} \Big (  ( \bar{P}^{-1}  +  C^T  R^{-1}  C   ) \notag  \\
       & \qquad  - ( \bar{P}^{-1}_{\sharp}  +  C^T  R^{-1}  C   ) \Big)   ( \bar{P}^{-1}  +  C^T  R^{-1}  C   )^{-1}   A_{\sharp}   \notag  \\
       = & A_{\sharp} ( \bar{P}^{-1}_{\sharp}  +  C^T  R^{-1}  C   )^{-1}  \Big ( \bar{P}^{-1} -  \bar{P}^{-1}_{\sharp}   \Big )   \notag  \\
       & \qquad \qquad \qquad  \times  ( \bar{P}^{-1}  +  C^T  R^{-1}  C   )^{-1}    A   \notag  \\
       = & A_{\sharp} ( \bar{P}^{-1}_{\sharp}  +  C^T  R^{-1}  C   )^{-1} \bar{P}^{-1}_{\sharp} \Big ( \bar{P}_{\sharp}  - \bar{P}    \Big )  \bar{P}^{-1}  \notag  \\
       & \qquad \qquad \qquad \times   ( \bar{P}^{-1}  +  C^T  R^{-1}  C   )^{-1}    A .   \notag
       \end{align}
   By denoting $
     \tilde{A}_{\sharp}  =  A_{\sharp} ( \bar{P}^{-1}_{\sharp}  +  C^T  R^{-1}  C   )^{-1} \bar{P}^{-1}_{\sharp}$ and $
     \tilde{A}  =  A ( \bar{P}^{-1}  +  C^T  R^{-1}  C   )^{-1} \bar{P}^{-1},$
   we have that $\bar{P}_{\sharp} -  \bar{P} $ in \dref{equ:errorbarP} satisfies
   \begin{align} \label{equ:errorPPP}
     \bar{P}_{\sharp} -  \bar{P} & \leq M_{1} \epsilon I_n + \tilde{A}_{\sharp} (\bar{P}_{\sharp} -  \bar{P}) \tilde{A}^T \notag \\
     & \leq M_{1} \epsilon I_n + M_{1} \epsilon \tilde{A}_{\sharp}   \tilde{A}^T + \tilde{A}_{\sharp}^2 (\bar{P}_{\sharp} -  \bar{P}) (\tilde{A}^T)^2 \notag \\
     &   \qquad \qquad \vdots \\
     & \leq M_{1} \epsilon \sum_{h=0}^{\infty}  \tilde{A}_{\sharp}^{h}   (\tilde{A}^T)^{h} + \tilde{A}_{\sharp}^{\infty} (\bar{P}_{\sharp} -  \bar{P}) (\tilde{A}^T)^{\infty}. \notag
   \end{align}
   To proceed, by expanding
   \begin{align}
     & (\bar{P}^{-1}  +  C^T  R^{-1}  C   )^{-1}   = \bar{P} - \bar{P}C^T (R + C \bar{P} C^T) C \bar{P}, \notag
   \end{align}
   we can find that
   \begin{align}
     \tilde{A}  = & A   -   A  \bar{P}  C^T (R + C \bar{P}  C^T) C  = A  ( I_n - L_{\infty} C),  \notag
    \end{align}
   is a closed-loop state matrix and Schur stable. Similarly, $\tilde{A}_{\sharp}$ is Schur stable.
   Hence, we have $\tilde{A}_{\sharp}^{\infty} (\bar{P}_{\sharp} -  \bar{P}) (\tilde{A}^T)^{\infty} = 0$ in \dref{equ:errorPPP}, and $
     \bar{P}_{\sharp} -  \bar{P} \leq M_{1} \epsilon \sum_{h=0}^{\infty}  \tilde{A}_{\sharp}^{h}   (\tilde{A}^T)^{h}.$
   According to \cite[Lemma 3, Theorem 2]{qian2023observation}, there exists a positive constant $M_2$ such that $\sum_{h=0}^{\infty}  \tilde{A}_{\sharp}^{h}   (\tilde{A}^T)^{h} \leq M_2 I_n,$
   which leads to $  \bar{P}_{\sharp} -  \bar{P} \leq M_{1} M_{2} \epsilon I_n.$
   By combining the above results in this proof and the fact $ \epsilon \sim \mathcal{O}   ( \sqrt{ 1/N} ) $,
   we have $  \| \bar{P}_{\sharp} - \bar{P}  \|_2  \leq   \mathcal{O} \big ( \sqrt{ {1}/{N }} \big),$
    when $N \ge N_0 (\epsilon,\delta)$ and $\epsilon < \epsilon_0$. Similarly, we have
  \begin{align} \label{equ:gappp}
      \|  {P}_{\sharp} -  {P}  \|_2 & \leq   \mathcal{O} \big ( \sqrt{ {1}/{N }} \big),
  \end{align}
holds with probability at least $1-\delta$, when $N \ge N_0 (\epsilon,\delta)$ and $\epsilon < \epsilon_0$ with $N_0(\epsilon,\delta)$ and $\epsilon_0$ being defined in \dref{equ:NG} and \dref{equ:epsilon0}, respectively. Next, it follows from \dref{equ:systemstate}  and \dref{equ:subml} that
  \begin{align}
   e_{k+1} = & \ (I_n - L^{\sharp} C ) \hat{A} e_k +  (I_n - L^{\sharp} C ) (\hat{A} - A) x_k \notag \\
   &  - (I_n - L^{\sharp} C ) \omega_k + L^{\sharp} \nu_{k+1}. \notag
  \end{align}
  Then, its covariance satisfies
  \begin{align}
   &  P_{e,k+1}
   =   (I_n - L^{\sharp} C ) \hat{A} \mathbb{E}\{e_{k}e_{k}^T\}  \hat{A}^T (I_n - L^{\sharp} C )^T \notag \\
   & + (I_n - L^{\sharp} C ) \hat{A} \mathbb{E}\{e_{k}x_{k}^T\} (\hat{A} - A)^T (I_n - L^{\sharp} C )^T \notag \\
   &  +  (I_n - L^{\sharp} C ) (\hat{A} - A) \mathbb{E}\{x_{k}e_{k}^T\}  \hat{A}^T (I_n - L^{\sharp} C )^T \notag \\
   &  +  (I_n - L^{\sharp} C ) (\hat{A} - A)  \mathbb{E}\{x_k x_k^T\} (\hat{A} - A)^T  (I_n  \notag \\
   &  - L^{\sharp} C )^T + (I_n - L^{\sharp} C ) Q (I_n - L^{\sharp} C )^T + L^{\sharp} R (L^{\sharp})^T \notag \\
   \leq &  (1 + \eta) (I_n - L^{\sharp} C ) \hat{A} P_{e,k} \hat{A}^T (I_n - L^{\sharp} C )^T +  L^{\sharp} R (L^{\sharp})^T \notag \\
   &  + (1 + 1/\eta)  (I_n - L^{\sharp} C ) (\hat{A}- A)  \mathbb{E}\{x_k x_k^T\} (\hat{A}- A)^T   \notag \\
   & \times (I_n - L^{\sharp} C )^T + (I_n - L^{\sharp} C ) Q (I_n - L^{\sharp} C )^T , \notag
  \end{align}
 where $\eta < \min \{\|\hat{A} - A\|_2, \  1/|\lambda ( (I_n - L^{\sharp} C ) \hat{A})|^2 - 1 \},$  and ``$\leq$'' is derived based on the fact that $AB^T + BA^T \leq \eta AA^T + 1/\eta BB^T$ holds for any square matrices $A$ and $B$, and any positive scalar $\eta$. According to Assumption \ref{xbound}, we  further have
 \begin{align} % \label{equ:peinequality}
   & P_{e,k+1}
   \leq  (1 + \eta) (I_n - L^{\sharp} C ) \hat{A} P_{e,k}  \hat{A}^T (I_n - L^{\sharp} C )^T + M_3,  \notag
  \end{align}
  where
  \begin{align}
   & M_3  = \ (1 + 1/\eta)  (I_n - L^{\sharp} C ) (\hat{A} - A)  \Pi (\hat{A} - A)^T     \notag \\
   &  \times (I_n - L^{\sharp} C )^T + (I_n - L^{\sharp} C ) Q (I_n - L^{\sharp} C )^T +  L^{\sharp} R (L^{\sharp})^T, \notag
  \end{align}
 is a constant matrix. Due to $ \eta < \min \{ 1/|\lambda ( (I_n - L^{\sharp} C ) \hat{A} )|^2 - 1 \},$ we can find that $\sqrt{ 1 + \eta} (I_n - L^{\sharp} C ) \hat{A}$ is Schur stable since $(I_n - L^{\sharp} C ) \hat{A}$ is Schur stable \cite{935054}. Hence, the above inequality can guarantee that $P_{e,k+1} $ is uniformly bounded. Next, the error between $P_{e,k} $ and $P_{\sharp}$  in \dref{equ:Kalmannewone} is analyzed. According to the above inequality and \dref{equ:Kalmannewone}, we can derive that
 \begin{align}
     P_{e,k+1} - P_{\sharp} \leq & (I_n - L^{\sharp} C ) \hat{A} (P_{e,k} - P_{\sharp})  \hat{A}^T (I_n - L^{\sharp} C )^T \notag \\
     &  + M_{4} \|\hat{A} - A\|_2 I_n, \notag
  \end{align}
  where $M_{4}$ is a positive constant scalar. Subsequently,
  \begin{align}
   & \ P_{e,k+1} - P_{\sharp} \notag \\
    \leq & \big ((I_n - L^{\sharp} C ) \hat{A} \big )^{k+1} (P_{e,0} - P_{\sharp})  \big (\hat{A}^T (I_n - L^{\sharp} C )^T  \big )^{k+1} \notag \\
    +  &  \|\hat{A} - A\|_2 \sum_{h=0}^{k} M_4  \big ((I_n - L^{\sharp} C ) \hat{A} \big )^{h}    \big ( \hat{A}^T (I_n - L^{\sharp} C )^T  \big )^{h} , \notag
 \end{align}
  When $k$ tends to infinity, we have $ P_{e,\infty} - P_{\sharp}  \leq M_5 \|\hat{A} - A\|_2,$ where $
    M_5   =   M_4 \sum_{h=0}^{\infty}   \big ((I_n - L^{\sharp} C ) \hat{A} \big )^{h}    \big ( \hat{A}^T (I_n - L^{\sharp} C )^T  \big )^{h}$
 is a constant matrix according to \cite[Theorem 2]{qian2023observation}. With the results in Theorem \ref{thm5} and \dref{equ:gappp},  there exists a positive constant $M_e$ that
 \begin{align}
  &  \| P_{e,\infty} - P  \|_2 \leq \|  P_{e,\infty} - P_{\sharp} \|_2 + \|  P - P_{\sharp} \|_2  \leq   \mathcal{O} \big ( \sqrt{ {1}/{N }} \big),  \notag
 \end{align}
 with probability at least $1 -  \delta$, when $N \ge N_0 (\epsilon,\delta)$ in \dref{equ:NG} with $\epsilon < \epsilon_0$. Till now, we have ensured Theorem \ref{thm7} for the case where $ \| \hat{A} - A \|_2 \neq 0$, while $ \| \hat{B} - B \|_2= \| \hat{C} - C \|_2= \| \hat{R} - R \|_2 = \| {P}_{\sharp} -  \hat{P}_{\sharp}  \|_2=0$.

 Next, let us consider the following cases one by one: \\
 Case 2): $ \| \hat{C} - C \|_2 \!=\!  \| \hat{R} - R \|_2 \!=\! \| {P}_{\sharp} -  \hat{P}_{\sharp}  \|_2 \!=\!  0$; \\
 Case 3): $  \| \hat{R} - R \|_2 \!=\! \| {P}_{\sharp} -  \hat{P}_{\sharp}  \|_2 \!=\!  0$; \\
 Case 4): $\| {P}_{\sharp} -  \hat{P}_{\sharp}  \|_2 \!=\!  0$; \\
 Case 5): The original case in Theorem \ref{thm7}.

To differentiate, we label $P_{e,\infty} $ in Cases 1--5 as $P_{e,\infty}^1 $, $P_{e,\infty}^2 $,$P_{e,\infty}^3 $, $P_{e,\infty}^4 $, and $P_{e,\infty}^5 $, respectively. Note that the errors $\| \hat{B} - B \|_2$, $\| \hat{C} - C \|_2$, $ \| \hat{R} - R \|_2$, and $\|  {P}_{\sharp} -  \hat{P}_{\sharp}  \|_2$ are bounded by $\mathcal{O}   ( \sqrt{ {1}/{N }}  )$, and all parameters in these cases are bounded. By leveraging the techniques about the Riccati equation used in Case 1, we have $ \| P_{e,\infty}^1 - P  \|_2$, $\| P_{e,\infty}^2 - P_{e,\infty}^1  \|_2$, $\| P_{e,\infty}^3 - P_{e,\infty}^2  \|_2$, $ \| P_{e,\infty}^4 - P_{e,\infty}^3  \|_2$, and $ \| P_{e,\infty}^5 - P_{e,\infty}^4  \|_2$
%\begin{align}
%& \| P_{e,\infty}^1 - P  \|_2, \ \| P_{e,\infty}^2 - P_{e,\infty}^1  \|_2,  \ \| P_{e,\infty}^3 - P_{e,\infty}^2  \|_2, \notag \\
%  & \| P_{e,\infty}^4 - P_{e,\infty}^3  \|_2, \ \| P_{e,\infty}^5 - P_{e,\infty}^4  \|_2,  \notag
%\end{align}
are upper bounded by $\mathcal{O}   ( \sqrt{ {1}/{N }}  )$
 with probability at least $1-\delta$, when $N \ge N_0 (\epsilon,\delta)$ and $\epsilon < \epsilon_0$. Hence, we have that $ \| P_{e,\infty}^5 - P  \|_2 \leq \mathcal{O} ( \sqrt{ {1}/{N }}  )$ holds with probability at least $1-\delta$, when $N \ge N_0 (\epsilon,\delta)$ and $\epsilon < \epsilon_0$. Thus, the proof of Theorem \ref{thm7} is complete.

\bibliographystyle{ieeetr}
\bibliography{ref}

%\begin{IEEEbiography}
%[{\includegraphics[width=1in,height=1.25in,clip,keepaspectratio]{Author-duanpeihu}}]
%{Peihu Duan}~received the B.Eng. degree in Mechanical Engineering from Huazhong University of Science and Technology, Wuhan, China, in 2015. He received the Ph.D. degree in Mechanical Systems and Control from Peking University, Beijing, China, in 2020. Currently, he works as a Postdoc at the Department of Electrical and Electronic Engineering in the University of Hong Kong, Hong Kong, China. From May 2019 to August 2019, he was a Research Assistant with the City University of Hong Kong, Hong Kong, China. From October 2020 to August 2021, he was a Postdoc at the Department of Electronic and Computer Engineering in the Hong Kong University of Science and Technology, Hong Kong, China. His research interests include cooperative control and state estimation of networked systems.
%\end{IEEEbiography}

\end{document}